\documentstyle[preprint,prl,aps]{revtex}

\tightenlines

\begin{document}

.

\vspace{5.0cm}

\begin{center}

\Huge

{\bf \underline{DID TIME BEGIN? WILL TIME END?}}

\normalsize

\vspace{4cm}

{\bf Paul H. Frampton}

University of North Carolina at Chapel Hill.

\end{center}

\newpage

.

\newpage
.

\vspace{4cm}

\LARGE

\begin{center}

\underline{\bf Table of Contents}

\end{center}

\normalsize

\vspace{2cm}

Preface \dotfill    5 

\bigskip
\bigskip

Chapter 1: Why do many other scientists believe time began at a Big Bang? \dotfill    7 

Chapter 2: Smoothness of the Universe.  \dotfill 16 

Chapter 3: Structure in the Universe  \dotfill  23 

Chapter 4: Dark Matter and Energy \dotfill  30 

Chapter 5: Composition of the Universe's Energy. \dotfill  36 

Chapter 6: Possible Futures of the Universe \dotfill  42 

Chapter 7: Advantages of Cyclic Cosmology \dotfill 48 

Chapter 8: Summary of Answers to the Questions: 

\indent ~~~~~~~~~~~~~~~ Did Time Begin? Will Time End? \dotfill  55 

\bigskip
\bigskip

Glossary \dotfill  56

\newpage

.

\newpage

\vspace{4cm}

\begin{center}

\LARGE

{\bf \underline{ Preface }}

\normalsize

\end{center}

Did time begin at a Big Bang?
Will the present expansion of the universe last for a finite
or infinite time? These questions sound philosophical
but are becoming, now in the twenty-first century, central
to the scientific study of cosmology. The answers,
which should become clarified in the next decade or two,
could have profound implications for how we see our own
role in the universe.  Since the original publication
of Stephen Hawking's {\it A Brief History of Time}
in 1988, the answers to these questions have 
progressed as a result of research by the community
of active theoretical physicists including myself. 
To present the underlying ideas requires discussion
of a wide range of topics in cosmology, especially the
make up of the energy content of the universe.
A brief summary of my conclusions, that of three different
possibilities concerning the history and future of time, the
least likely is the conventional wisdom (time began and will never end)
and most likely is a cyclic model (time never begins or ends),
is in the short final Chapter which could be read 
first. To understand the reasoning leading
to my conclusions 
could encourage reading of my entire book.
My hope in writing this, my first popular book,
is that it will engender reflection about time. Many a non-scientist 
may already hold a philosophical opinion about
whether time begins and ends. This book's aim
is to present some recently discovered
scientific facts which can focus the reader's consideration of
the two brief questions in my title.

\vspace{1cm}

\hspace{12cm} Paul H. Frampton

\hspace{12cm} Chapel Hill

\newpage

.
\newpage

\bigskip
\bigskip

\LARGE

\begin{center}

{\bf \underline{ Chapter 1}}

{\bf WHY DO MANY OTHER SCIENTISTS BELIEVE TIME BEGAN AT A BIG BANG?}

\end{center}

\normalsize

\bigskip
\bigskip
\bigskip

Our everyday perception of the universe comes
from looking up at the sky to see the Sun in the daytime
and, more particularly, to see thousands of stars 
in the night-sky. 
Surely some of the oldest questions since the beginnings of human thought
are: How large is the universe? Did it ever begin? 
What are the principal constituents of the present
universe? Will time ever end?

Cosmology is the name for the scientific study of the
universe. The present time is an unprecedented age for
cosmology because it is fair to say that in the last five years
we have learned more in cosmology than in all previous
human history. 
Despite this enormous and exciting growth of our knowledge
as a result of many impressive observations, the universe has
become more enigmatic in many ways. The more we learn, the more 
the extent of our ignorance becomes manifest.

Cosmology has recently answered some of the old questions and in
this Chapter we shall give answers to the first two:
How large is the universe? How long ago did it, or at least
the present expansion era, begin? We can all agree that the expansion stage
we are presently in began a finite time ago but, as I shall
explain later in ths Chapter, it
is not obvious that time itself began then, if ever. 

We do know how large the visible universe is, meaning how
far away are the most distant galaxies whose light can have 
reached us on the Earth. It is theoretical possible, 
and even favored in
some theoretical scenarios, that our universe is actually
very much larger, than the visible universe. 
In some very speculative scenarios the universe is spatially finite
with non-trivial topology. This is at present 
not readily testable so we shall be content to
try to convey just how gigantic is the visible part.

The observational means by which we know accurately
the size of the universe will not concern us here but
sufficient to say that present studies using the Hubble Space
Telescope combined with the largest (up to 10 meters in
diameter) ground-based optical telescopes tell us the size
of the visible galaxy to an accuracy of a few per cent. This
sort of accuracy has been achieved only since the turn of the 21st
century.

\bigskip
\bigskip

Cosmological distances are so much bigger than any distance
with which we may be familiar that it is not easy to grasp
or comprehend them even in our imagination. So let us begin with
the largest distance which is easily comprehensible from
the viewpoint of our experience.

A very long airplane ride may take 15 hours and go 9000 miles,
a significant part of half way around 
the Earth. People who travel a lot like some physicists
may take such a flight a few times each year. One knows
that the plane has a ground speed of about 600 miles per
hour and the discomfort of sitting, especially in economy
class, for such a long time gives a strong impression of
just how far that distance is. Of course, people a hundred 
years ago would never travel that far in a day but now we do
and it gives us a feel for the size of the planet
so that is a length distance from which we can begin.

The next larger distance to think 
about is the distance between the Earth and the Moon. 
This is about thirty times the distance of the plane ride
and so would take some three weeks at the same airplane speed,
a few days in a NASA spacecraft. The distance to the Moon
is thus imaginable: if you walked at four miles an hour
non-stop without sleep it would take about eight 
years to arrive and another eight years to return.
Nevertheless, the arrival on the Moon of astronauts
Armstrong and Aldrin in July 1969
was one of the most memorable events of the last century.
Only partly was it due to the distance to the Moon, it
was equally the concept of humans
walking for the first time
on an astronomical object other than the Earth.

The Moon is visible in the night sky, and equally often
in the daytime thanks to its reflection of light from
the Sun. The Sun is by far our nearest star and its radiated
energy is crucial to the possibility of life on Earth.
How far away is the Sun? It is about ten thousand times the
length of the airplane ride and would take about twenty
years to reach at the speed of the airplane. Not 
that any sane person would want
to go there with a surface
temperature well above that of
molten iron. The Sun is about four hundred times
further away than the Moon, and is already at such a large distance
that it far exceeds anything with which we are familiar. This
sets the scale of the Solar System with the Earth,
rotating on its axis once a day, orbiting once
a year around the Sun at a distance of some ninety-three
million miles. Other planets, Mercury and Venus, circulate
inside the Earth's orbit while six others including Mars, 
Jupiter and Saturn
orbit outside the Earth.

It is almost inconceivable that any human being will travel 
outside of the Solar System in our lifetimes just because
of its enormous size. Yet on the scale of the visible universe the
Solar System is, in contrast, unimaginably tiny and insignificant.
So if there were no life other than on the Earth
the universe
would seem to be an almost absurdly large object if life were its
primary goal. 

In addition to the Moon and some planets,
we can see thousands of stars
with the naked eye. 
Most of these stars are similar to our Sun but appear
much dimmer because of their distance. How
distant are even the nearest stars? The answer is some 
two million times the distance to the Sun. So whereas
we can reach the Sun in twenty years at the speed of an airplane
to reach the nearest star in twenty years would require
a two million times faster speed. A quick
calculation shows this takes six hundred miles per hour
into thirty-five thousand miles per second. To put
such a speed into perspective, the speed of light
is about one hundred and eighty thousand miles
per second. This means our imaginary airplane,
suitably coverted as a spacecraft, must
travel at one fifth of the speed of light
just to reach the nearest star in twenty years.

Here we see the limitations to any travel possibilities
not only in our lifetime but what would seem to be for ever.
According to the theory of relativity, which there is no reason to
doubt, nothing can travel faster than the speed of light. So even
if the human lifetime is extended by medical advance to
two hundred years or even a thousand years it is impossible
to travel during one lifetime more than a few hundred
times the distance to the nearest star. But the
galaxy to which our Solar System belongs extends about
ten thousand times the distance to the nearest star.
So it would seem impossible ever to leave our particular galaxy
which is known as the Milky Way from its appearance spreading 
across the night sky.

There are a couple of holes in this argument. First, according to relativity
time slows down as one travels at approaching the speed of light.
Second, it is conceivable that some cryogenic method might
be devised to slow down the speed of life and greatly 
enhance the effective human lifetime.
Even so, to travel outside our galaxy does seem forever impossible and
cosmology may remain a spectator sport.

One hundred years ago it was generally believed that the 
the universe was comprised of only the Milky Way. 
The size of the our galaxy is 
only ten thousand times the distance to the nearest star which is itself
two hundred thousand times the distance to the Sun. That is, the
galaxy size is two billion times (one billion is a thousand million)
the Earth-Sun distance. The size of the galaxy seems to be relatively
independent of time and so in ignorance of a universe very much bigger than
a single galaxy it was believed,
before the 1920s, that the universe was itself static,
neither expanding nor contracting.

When the general theory of relativity was proposed in 1915 this state
of the observational knowledge stymied what could have been predicted,
namely the overall expansion of the universe. This expansion, which
is a key feature of the universe and will lead us to the conclusion that
it had a definite beginning, became an option only by observations somewhat
later in the 1920s.

Now we arrive at the final leap in the distance scale. The visible universe
turns out to be about four hundred thousand times 
the size of the Milky Way, very much larger
than previously imagined. That is,
not only is the Solar System of neglible size with respect to the universe
but so is even the entire Milky Way. In fact, in theoretical
cosmology galaxies are treated as point particles! And
the human race may be confined forever to be inside
one of these points!

We have seen that the size of the galaxy is tremendously bigger, by a factor
of billions, than the distance to the Sun. Yet the visible universe
is so much larger again than a galaxy that to study it
each galaxy may be regarded as just a single dot within it.
This should communicate an idea in words of just how big the visible 
universe is. Now we show how
we know the present expansion (and
possibly time itself) had a beginning some fourteen billion years ago.

\bigskip
\bigskip

As already discussed, the size of the Milky Way has not expanded or contracted
significantly since it was formed some ten billion years ago. Within
the Milky Way the Sun and the Solar System appeared about five billion years ago.
The Earth is a little younger, about four and a half billion years. 
The point is the general arrangement of the Sun and planets in the Solar System
has not markedly changed in the last few billion years. During that time
we may regard the galaxy and its contents as of a constant size.

A truly astonishing revelation comes when we study the same question
for the entire universe including hundreds of thousands of the the billions
of galaxies outside of the Milky Way. The issue is what is their typical
motion relative to our galaxy?

Here it is important to understand a phenomenon well-known in physics called
the Doppler effect. It is a more
familiar phenomenon for sound waves than for light. When a train blows its
whistle and passes a listener the pitch of the whistle falls from a higher
note to a lower note. In fact not only the whistle but the entire train noise
exhibits the same Doppler effect. Why does this happen? It is because the 
motion of the train towards the listener compresses the sound waves to a 
shorter wavelength and, because the velocity of sound is unaltered, 
to a higher frequency. Similarly, when the train is moving away the sound 
waves are stretched and the frequency lowers. The pitch for a stationary
train would lie between the two pitches while approaching and receding.
The shift in frequency is calculable simply in terms
of the ratio of the speed of the train and the speed of sound.

\bigskip
\bigskip

Exactly the same Doppler effect occurs for light. If a galaxy
is approaching our galaxy, its light
appears with a higher frequency. For the
visible spectrum the highest frequency is for blue light
so we may say that the light is blue-shifted. On the other hand,
if the galaxy is receding from ours its light appears shifted
to a lower frequency and is red-shifted toward the red or lowest
frequency end of the visible spectrum.

In fact what are observed are the spectral lines of light
emitted from known atoms and whose frequencies are accurately known 
here on Earth. If all the lines are systematically 
shifted towards the blue
then the galaxy hosting these atoms is approaching 
the Milky Way: if toward the red then it 
is receding from us. This can be made precise 
by a mathematical expression for the frequency
shift which gives the approach or recession speed as a fraction
of the speed of light.

When this is studied for a large number of galaxies it might
be expected that roughly half would be blue-shifted and half
red-shifted if the overall universe were
static and the galaxies were moving randomly.

What was observed, however, to the astonishment of Hubble and
Einstein in 1929 is that almost all galaxies are red-shifted.
Apart from a few galaxies in our immediate neighborhood
like the nearby Andromeda galaxy 
all the hundreds of thousands of more distant 
galaxies measured are receding from us.
This means that the entire universe is expanding and the galaxies 
are moving away from us and as we shall see later, from each other.
This phenomenon is called the Hubble expansion.

The next important question is how does the recession
velocity depend on the distance of the galaxy from the Milky Way?
The galaxies can be classified into types such as spiral, elliptical,
irregular and the total light emitted may be assumed to show
regularity within each type. But the apparent brightness on Earth depends
on the distance and falls off as an inverse square law. So the apparent
brightness can be converted into a distance.
When the recession velocity is compared to the distance 
a very important regularity appears. This is the most significant 
discovery in cosmology
called Hubble's law which states that the recession velocity is proportional to
the distance. The ratio of the velocity and the distance
is thus a constant, the Hubble parameter. Its value is notoriously difficult
to measure but now we do know it to be close to seventy, within ten percent,
in certain units. The units which are not crucial to the general discussion
are kilometers per second per megaparsec where a megaparsec is the
distance light travels in about three million years.

How does this tell us when the present expansion phase
began? This requires the use of the 
equations of the general theory of relativity together with
two assumptions. The first assumption is that 
the universe at the large scale is the same,
on average, in each of the three directions of space. This is called isotropy
and is supported by observations which indicate no preferred direction
exists in the universe. The second assumption is that, on average,
all positions in the universe are equivalent. This
means that in all galaxies (hypothetical!)
observers would see the recession of other galaxies according to
Hubble's law. This assumption is called
homogeneity. The combination of these two assumptions, isotropy and homogeneity,
is known technically as the cosmological principle.

Combining the cosmological principle with the general theory of relativity
gives rise to mathematical equation known after its inventor
as the Friedmann equation
which characterizes the expansion of the universe in terms of a scale 
factor which is a function of time and specifies the typical
distance between galaxies. Inserting the known Hubble parameter 
and the present known composition
of the universe then enables us to calculate the scale for all past time.
Note, in passing, that we cannot do this for future times with confidence
because we do not know with certainty
how the composition of the universe will
evolve in the future.
For the past we have good confidence and we find a striking conclusion:
run in reverse, the contracting universe is seen
mathematically to shrink to a point at a well-defined
past time. At that time the universe may have begun
in some unimaginably
powerful explosion called the Big Bang. Our present expansion
 seems to have a beginning and we know when.
It was some 13.7 billion years ago, give or take a two hundred million
years.
The age of the universe is now established to an accuracy of about two percent.

\bigskip
\bigskip

There is one serious problem with extrapolating the Friedmann'equation
back in time, namely that about 13.7 billion years ago
the equation becomes singular. The density and temperature
become infinite and the classical theory
of general relativity breaks down. Thus the Big Bang, only much
more recently named, was the {\it initial singularity}
to the early workers of the 1920s and 1930s.

Within the theory therefore we know that the Big Bang must be avoided.
A common response is to invoke quantum mechanics. From the
fundamental constants, the speed of light,
Newton's gravitational constant and Planck's constant
one can construct a time known as the Planck time
which is a tiny fraction ($10^{-44}$)
of a second. One may say that at that short time
after the would-be Big Bang the classical
theory of general relativity must break
down. Quantum gravity must play a role and no completely
satisfactory theory exists. So one may argue that quantum mechanics
rescues the day. Indeed an entire filed
known as quantum cosmology has been built up
around such a ruse but without really new insights. 

In this book we shall not appeal to quantum mechanics
in this way but examine whether the Big Bang can be avoided
in a purely classical context.

\bigskip
\bigskip

Before proceeding there is one amusing anecdote about the origin of the
graphic term Big Bang which seems so apt to describe the beginning of the universe.
Before the scenario we have just described was firmly established 
a competing theory was the steady-state theory which postulated that despite the Hubble
expansion there was a steady state and no beginning
because of the continuous spontaneous creation of new galaxies.
As a derogatory term for the competing theory, Big Bang was coined by
a leading exponent of the steady-state theory. Unfortunately for him, the Big Bang
theory and not his viewpoint was confirmed by subsequent measurements.

\bigskip
\bigskip

There is one alternative view 
(we shall discuss another view in Chapter 7)
of the Big Bang where the lifetime 
of the (generalized) universe
is infinite. The process of the Big Bang is in that view something
which has occurred repeatedly, indeed infinitely, resulting in an
infinite number of different universes of which we are aware of just one.
This is technically called eternal inflation and the resultant
universe becomes a multiverse for obvious reasons.

It is difficult, if not impossible, to test eternal inflation because
the other universes would seem to be forever hidden from our view.
The best chance may be to make a probabilistic treatment of the multiverse
to estimate the probability of the universe we observe having the 
properties it has in terms of its fundamental consituents or building blocks.
Some research is indeed being pursued along this line.

\bigskip
\bigskip

In this book we shall assume the 
beginning of the present expansion era to have taken place
approximately fourteen billion years ago followed,
as will be discussed, by an inflationary era
of rapid expansion. The latter explains two different kinds
of extraordinary smoothness
observed. We know that there was temperature uniformity to
one part in one hundred thousand in the universe when it was only
four thousand years old.
Then there is the proximity of the observed density 
of the universe to a special value known as the critical density
which would, without inflation, require preternatural fine-tuning
in the early universe.

Inflation appears now to be ubiquitous in almost all theoretical
cosmology, in one form or another. As we shall discuss, it can account
for the otherwise-puzzling smoothness properties of the universe.
On the other hand, it is exceedingly
difficult to make direct measurements which are sensitive
to such an early era, the inflationary era 
which occurred even earlier than a billionth
of a second after the would-be Big Bang. 

Normal observations involving electromagnetic radiation
go back only to a few
hundred thousand years after the would-be Big Bang, far too
recent to study inflation. Studies of abundances of light elements
like helium and hydrogen probe indirectly back to a cosmic time
of one second after the would-be Big Bang. Potential neutrino 
astronomy measurements could directly
probe a similar era.

The only chance of direct observation of the inflationary era 
would appear to be by gravitational radiation - waves created by
string gravitational fields in the early universe.
The observability of such radiation depends on how early inflation
took place, the earlier being the easier to detect.
For later inflation it looks presently impossible to detect such gravity waves.
The word "presently" is essential since how technology will evolve, 
and what consequent scientific apparatus will be enabled, 
by the end of the 21st century is impossible to predict.
It is a lesson from the history of physics that  
to decree anything impossible is a dangerous prediction.  

\bigskip 
\bigskip 

One may ask what happened before the Big Bang, if it did occur? 
This is beyond scientific investigation and it is easier 
to assume that time began then. Very ambitious and speculative 
theories discuss prior times using ideas such as
T-duality in string theory or eternal inflation with its resultant
multiverse. If such theories become testable and shed light
on the physics of our universe then they
must be taken very seriously in a more general
domain of applicability. At present, such ideas remain pure 
speculation.

Another question which we shall address at length 
in this book is what will happen to the universe in the future?
This is understood less than the past, and depends critically on the
properties of the newly-discovered Dark Energy which comprises almost
three-quarters of the total energy density of the universe.

\bigskip
\bigskip

Concerning space one will ask whether it too, like past time, can
be finite in extent. Is it possible that by proceeding in a straight
line one will return, after a finite time and distance, to the starting
point due to a non-trivial topology of space? There is
no compelling evidence for this possibility though certain
data on the cosmic microwave background radiation can be interpreted
as supporting such an assumption. Alternative explanations for the
data come from arguments about cosmic variance or from
small distortions in the hypothetical inflaton potential, so
the case for non-trivial spatial topology is not at all strong.
If there were non-trivial topology, it could be of one of three types.
Positive curvature corresponds to a closed universe, negative
curvature to an open universe and flat, without curvature, 
is the geometry predicted by inflation. The local properties
in such a universe satisfy the same general relativity equations
as for the case of infinite space with trivial topology;
only the global toplogical properties differ so it is
not obvious from, say, study of our galaxy alone which option
Nature chooses. The notion of non-trivial topology of space
necessarily introduces at least one fundamental length
which, to be consistent with observational data, must be
comparable to the radius of the visible horizon
of a few gigaparsecs.

A common question by an educated non-physicist is into
what does the universe expand or, equivalently,
what is "outside" the universe? So let us try to give a 
clear answer. The answer is not obvious only because of the
limitations to the human imagination. All of us can easily
imagine three spatial dimensions but four is enormously
more difficult. Unfortunately
the spacetime manifold of the universe is itself four
dimensional and this is both why the question naturally
arises and why the answer is slightly elusive. If we scale down
by one dimension there is an analogous situation which
is, by contrast, very easy to grasp. Take a balloon with spots on the surface
to represent galaxies. As time passes we inflate the balloon
and the spots get further apart as for the expansion of the universe.
Now a two-dimensional being on the balloon surface may 
ask: into what is this two dimensional space expanding?
The answer is that there is nothing "outside" the two-dimensional
surface as obviously it is a closed surface without boundary.
Similarly the three-dimensional space of our universe
has no boundary and no "outside".

\bigskip
\bigskip

Among so many interesting yet unanswered questions, the one about
the beginning of the present expansion 13.7 billion
years ago seems settled. The expansion itself was universally accepted only
in 1965 as a result of the discovery of the remnant background
microwave radiation. The uncertainty of the future scenario for the universe
is under much study as a result of the discovery of dark energy
dating from 1998. Thus we are at a very exciting time in the subject.

The establishment of a finite length of the present expansion of about fourteen billion
years is clearly of fundamental importance which could be equated
by establishment of a finite spatial extent to the universe.
There is absolutely no compelling evidence for such an idea although some
data from the WMAP analysis of the cosmic background radiation,
particularly the unexpectedly small values of the low
multipoles, has been interpreted as suggestive of finite size
and non-trivial topology.

Certainly these cosmological considerations change our picture
of our own history.

Finally, in our discussion of the universe's longevity,
it is important that we use a linear time, rather than logarithmic time,
in the above discussion. The two are dramatically different. Firstly in
logarithmic time the age becomes infinite. But the difference can be better
seen in a concrete analogy. 

Suppose we condense the entire cosmic history of fourteen billion years
into one day of twenty-four hours starting and ending at midnight.
First we use linear time. The nucleosynthesis takes place just
a trillionth of a second after midnight; recombination and the surface
of last scatter are three seconds later; galaxy formation starts
around 1.40am; the Sun is created about 4pm in the afternoon
and Julius Caesar invades Gaul about a hundredth of a second before
midnight.

But if we map the same history using logarithmic time starting at
the Planck time (since we must now start at a finite time in the past)
then the occurrence of major events looks completely different.
Nucleosynthesis waits until 5pm in the afternoon;
recombination and the surface of last scatter are at 10pm in
the evening; galaxy formation begins at 11.25pm;
the Sun is created at 11.48pm and Julius Caesar appears now only
a trillionth of a second before midnight.

This illustrates how the use of linear time in cosmology
effects the relative spacing of subsequent events. From the viewpoint of
fundamental physics more happened in the first second of the Big Bang
than has happened in the subsequent fourteen billion years,
more in parallel with a logarithmic picture of time.
But it is in linear time, with which we are familiar in measuring
all everyday events, that the time since the would-be
Big Bang does have the finite value of 13.7 billion years.

\bigskip
\bigskip

The answer to the question in this Chapter's title has already been alluded to,
that many other scientists (if not this author)
explain away the initial singularity of the Friedmann equation by
an appeal to quantum mechanics and quantum gravity.
The absence of a fully satisfactory theory of quantum gravity
can act as a further security for such scientists as no one can
definitively refute the argument.

However, the tentative attempts
at quantum gravity has problems.
The concept of the wave function of the universe, as
employed in quantum cosmology, is problematic for several reasons,
not least of which is that the observer is inside the system.
The Planck time is much shorter than the time expected
to pass between the would-be Big Bang
and the onset of inflation. Explaining away the initial singularity
by quantum mechanical arguments was useful only when it
was {\it faute de mieux}. This is no longer the case.

I find it more satisfactory to make a cosmological model where
the density and temperature are never infinite. This precludes
a Big Bang and replaces it with a different picture of time
where time never begins and never ends. This is in contrast with
the standard cosmological model where time begins at the Big Bang
and never ends during an infinite future expansion.

While I cannot prove rigorously that the conventional wisdom
is wrong, it does entail
the singularities and concomitant
breakdown of general relativity that we have mentioned.
The existence of plausible alternatives now, however, makes
the Big Bang idea less plausible.

As we shall show later in the book, there is an alternative
version where time begins at a Big Bang and ends in a "Big Rip"
at a finite time in the future. I regard this as prefrbale aesthetically
to the conventional picture. But best of three possibilities about time
is the "infinite in both directions", past and future" as
exemplified by a cyclic model My student and I constructed only in the twenty-first
century based on the dark energy component. Obseravations
of dark energy and its properties especially its equation of state
will confirm or refute such more satisfactory ideas about time which
insist there was never a Big Bang.

\newpage

\bigskip
\bigskip

\LARGE

\begin{center}

{\bf \underline{ Chapter 2}}  

{\bf SMOOTHNESS OF THE UNIVERSE}

\end{center}

\normalsize

\bigskip
\bigskip

When we look further at the universe in the large, with the galaxies
treated as point particles, there are even more surprises in store
beyond those already discussed which include 
the gigantic size, the cosmic expansion and the fact that the present 
expansion phase had a beginning.

What we have discussed already seems consistent however not obvious
to the naked eye. From
looking at the night sky we might most naturally suppose, as people did
for many hundreds of years before the 20th century, that the stars
had been there forever and would remain so. With the naked eye no further
progress could ever have been made. Powerful telescopes see objects,
galaxies, far more distant and completely outside the Milky Way. They are moving
away at high speed, higher as their distance increases. Ironically the only
other galaxy visible to the naked eye outside the Milky Way, the Andromeda
galaxy, happens to be moving towards us! But almost all the other galaxies are receding
from us.

To discuss the unexpectedly high smoothness of the universe, technically
called the horizon problem, we need to introduce the concept of temperature.
The universe is filled with radiation, electromagnetic radiation, which
is currently extremely cold. It is at a temperature of about 
three degrees above
absolute zero. The value of this temperature varies inversely as the 
scale factor which characterizes the size of the universe. 
Consequently as the universe expands the temperature is falling
even lower. Turning that around, if we ask instead about the past 
the temperature was higher. 

In the history of the universe, at least as far as the electromagnetic radiation
is concerned, a most important thing happened 
about three hundred thousand years
after the would-be Big Bang at which time the visible universe was about one thousand
times smaller than it is now. At that time the temperature was therefore one thousand
times higher than now. This means it was very hot, at three thousand degrees.
It turns out this is the maximum temperature below 
which hydrogen atoms can survive as bound states.
A hydrogen atom consists of a proton which forms the hydrogen nucleus
around which one electron orbits. At temperatures above three thousand degrees
the atom has a high propensity to ionize into a separate proton and electron.
Indeed at all times before this special 
"recombination" time the protons and electrons
existed separately in what is called an ionized plasma. After the special time
the hydrogen atoms existed as bound states. Technically this special
occurrence is illogically but for ever called recombination. The name is
illogical, it should be just combination, because the protons and electrons had
never previously been combined!

Electromagnetic radiation is composed of massless elementary particles called
photons travelling at the speed of light. Photons are scattered by charged
particles but not by neutral ones. This is why the recombination occurrence
is so important for the electromagnetic radiation in the universe. Before
recombination there was an ionised plasma of electrons and protons.
Such charged particles can scatter the photons. This means that the universe was opaque
until recombination. Photons
could not travel in straight lines at the speed of light because of
scattering by charged particles in the plasma. After
recombination, in contrast, the charged electrons and protons became bound into neutral
hydrogen atoms. The universe became transparent as it allowed
photons to propagate freely through it in straight lines since
neutral atoms have no interactions with photons.

\bigskip
\bigskip

This means that the photons detected from the universe as a whole, technically called
the cosmic microwave background radiation, have been able to travel along straight lines
at the speed of light for the full fourteen billion
years since the recombination occurred. This provides us with a unique and extremely valuable
means of studying the state of the universe just three hundred thousand years
after the beginning of the present expansion phase, the would-be
Big Bang. Much of the recent progress in cosmology
is due to the vastly improved experimental information 
since 1992 on the details of this background radiation.

At a temperature of three degrees above absolute zero the wavelengh of these
relic photons is about ten centimeters or four inches. This is in the 
part of the electromagnetic
spectrum called microwave and is of similar wavelength to the radiation
used in a domestic microwave oven. When they started out at recombination,
from what is technically known as the surface of last scatter, their wavelength
was one thousand times smaller or a tenth of a millimeter. This is because
the wavelength of a photon goes inversely as its energy. Energy is proportional to
temperature and the temperature was at that time one thousand times higher than now.
Another equivalent viewpoint is that the wavelength of the photon has been
stretched by a factor of a thousand, like everything else, during its
propagation to Earth from the surface of last scatter.

The background microwaves are invisible to the eye which can detect or see only
a narrow part of the electromagnetic spectrum in the visible region where
the wavelength is about a hundred thousand times smaller. But sensitive detectors
of the microwave radiation have been flown on satellites and in the upper atmosphere
suspended from large helium balloons. These detectors are looking out far beyond
all the galaxies to the surface of last scatter which represents the most distant
surface ever visible by electromagnetic radiation. To see further will require either
neutrinos or gravitational radiation but such more exotic observations are reckoned to
be extremely difficult. 

To explain why there is something extremely
remarkable about the observed smoothness of the universe at the surface of last scatter requires
that we take a close look at relativity theory. One prediction of relativity theory
is that nothing can move at a speed greater than that of light.
This applies not only to particles but also to any type of information.
In particular, a causative  phenomenon may not
create any effect faster than the speed of light. We are therefore influenced 
only by events in the earlier universe that  are such as can transmit information to us. 
Technically in special relativity
theory we say that an event can be affected only by earlier events
that are in its backward light cone. This requirement is called causality
and is extremely restrictive on possible physical theories. Relativity theory
has passed many tests and there is every reason to believe in it 
and in this consequent requirement of causality.

When the microwave radiation coming from the surface of last scatter is
analysed it is found that its temperature distribution is astonishingly
smooth. The temperature is exactly the same in all directions to an
accuracy of one in a hundred thousand. At that level there are exceedingly tiny fluctuations
which are themselves extremely important in the formation
of structure. But the important observation for the present discussion
is that the distribution
of temperature over this most distant surface is very, very smooth.
Why is that so surprising?

It is surprising because in the simplest Big Bang picture it would violate the sacred
principle of causality. Why is that? Because the diameter of the present visible universe
is roughly thirty billion light years and hence the diameter at the time
these photons were emitted  was one thousand times smaller or about thirty
million light years. A light year is the distance light travels in one year.
The age of the universe at recombination was just three hundred thousand years
so information could have travelled only three hundred thousand light years.
The radius of the
surface of last scatter is a hundred times bigger. So on the two-dimensional surface
there are the square of one hundred,
namely ten thousand, regions of the surface which
have never been in causal contact in the simplest Big Bang scenario.
So therefore it is amazingly enigmatic that the temperature
distribution is uniform within all these ten thousand
causally-disconnected regions to an accuracy of one part
in a hundred thousand!
This is the first example of smoothness.

It is very important to understand the previous paragraph because it
suggests that the Big Bang theory has to be modified. Before
describing the most popular modification of the Big Bang which accommodates
this feature, we shall describe a second example of smoothness which
will also be accommodated in the same modification, called inflation, to be
described at the end of this chapter.

\bigskip
\bigskip

The second example of smoothness observed for the
present universe is different but has a similarity with the first
that it is a property of the present universe which, when we extrapolate
back in time, seems utterly perplexing in the context of
Big Bang theory.

As already discussed in Chapter 1 the mathematical underpinning of the expanding
universe is provided by the general theory of relativity combined with
the assumptions of isotropy and homogeneity which make up
the cosmological principle. This leads to a simple differential equation
for the time dependence of the scale factor.

\bigskip
\bigskip

This equation tells us that the ultimate fate of the universe would
appear to depend on the overall density of energy and matter in
the universe. Of course we are talking only about the averaged density
and not a local density where there is structure such as galaxies and stars.
There is a critical value for this density which is such that the
universe would expand forever but more and more slowly such that it
will slow to a stop after an infinitely long time. With this critical
density another viewpoint is that the positive kinetic energy of the motions
of the galaxies and other matter in the universe exactly
balances the gravitational energy associated with the gravitational attraction.
Energy is conserved so the natural endpoint is where the galaxies come
to rest at infinite separations and therefore no residual
gravitational attraction.

If the density is less than this critical density the universe's geometry
has negative curvature and it will expand for ever without coming to a halt.
This negative curvature universe is technically termed open. The
final possible
case where the density is larger than the critical density has positive curvature.
In it the expansion will eventually halt followed by a recontraction to
a Big Crunch. In other words, for such a closed universe the 
gravitational energy more than overcomes the kinetic energy.
These simple arguments ignore the complications introduced by dark energy.
We shall return to this issue.

The special case where the density equals the critical density is at the borderline
between open and closed and is called flat. The three different geometries can
be characterized by whether the angles of a triangle add to less or more than, or
equal to the canonical one hundred and eighty degrees. This is the total
of the three angles in flat geometry, also called Euclidean geometry the
simplest form of geometry taught at school and very well known since the times
of the Greek civilization. The other two cases are examples
of non-Euclidean geometries,
discovered by mathematicians in the 19th century.
An example of positive curvature is provided by the surface of
a sphere such as the Earth. Drawing a triangle with one vertex at the North
pole and the other two vertices
on the equator a quarter of a circle from each other gives a
triangle that has all three angles equal to ninety degrees 
and a total of two hundred and seventy degrees.
The larger total characterizes a positive curvature as in a closed universe.
In a negative curvature as in an open universe the three angles
of a triangle sum to less than one hundred and eighty degrees.

The present total density of mass and energy
is equal to the critical density with an accuracy of a few percent.
In other words, the angles in any triangle really add up to
one hundred and eighty degrees to good accuracy.
This fact is very surprising in the Big Bang theory and suggests that
a component is missing. Why is that?

\bigskip
\bigskip

In the equations which describe the evolution of the scale factor
the flat geometry is an unstable solution. This means that if the
universe is flat now it most probably
was always exactly flat.
Deviations from flatness either in the open or closed direction
become rapidly magnified as the universe expands. To be so close
to flat now the density one second after the Big Bang must
have equalled the critical value to one part in a hundred trillion.

\bigskip
\bigskip

We have described two different forms of extreme smoothness exhibited
by our present universe in the large: the uniformity of the temperature
over the surface of last scatter to an accuracy of one part in a hundred thousand despite the
surface being comprised of ten thousand causally disconnected parts;
and the fact that the geometry is flat after such a long evolution of 13.7 billion years.
A priori both of these circumstances are exceptionally unlikely. 
This latter unlikeliness can be compared to balancing a pencil
on its sharpened point and finding it still so balanced a very long time later:
any physicist or non-physicist would demand an explanation. 

The appropriate modification of the Big Bang model is suggested by study of theories
for particle physics at high energies. Such theories involve dramatic transitions
between different phases as the temperature falls in the early universe.
As the universe cools it is possible for it to become temporarily
trapped in a state with peculiar properties called the wrong vacuum.
In the short time that it spends in this wrong vacuum or
ground state it is normal that the universe undergoes a very rapid
exponential expansion. To lead to adequate smoothness it is important
that the rapid expension be by at least twenty-eight orders of magnitude
in scale. This assumed period of super-fast expansion is called inflation.

It offers an explanation of both of the smoothness properties, the
horizon problem and the flatness problem, in one fell swoop. The
horizon problem is solved because the entire present visible universe
arises through inflation from just one tiny causally
connected region. The flatness is explained because the inflation
renders any pre-existing curvature negligible. This is like
taking a balloon and inflating it (without bursting!) to
an extraordinarily large size. The surface which at the beginning was
strongly curved would become essentially flat at the end of the process.

Indeed the only way known to accommodate the two shortcomings
associated with the smoothness properties of the Big Bang scenario
is to assume an inflationary era in the very early universe. There
is no direct evidence for inflation but the idea 
is regarded as likely
to survive in some form or other and provides an important new
ingredient in the Big Bang theory which became
established in the 1960s.

One of the most intriguing aspects of inflation is 
that it provides a possible mechanism
which disrupts the total smoothness initially 
by only exceptionally tiny perturbations.
This is an effect of quantum mechanics during the inflationary era
in the first billionth of a second after the would-be Big Bang.
These quantum fluctuations exit from the comoving Hubble radius,
characterizing the distance which is causally connected
during the inflationary era, and can re-enter very 
much later after some hundreds of thousands of years or more.
The question is whether these quantum fluctuations
can lead to the later perturbations needed to
seed the large scale structure of the universe.
This requires our understanding, more clearly than at present,
the connection between the quantum
fluctuations which exit the horizon
and the resultant classical perturbations seen in the study
of the cosmic background microwave radiation
from the surface of last scatter.
Nevertheless, it is a seductive idea
that gigantic structures such as clusters of
galaxies can be regarded as enormous
amplifications of once ultra-microscopic quantum effects.

\bigskip
\bigskip

Without inflation the extreme homogeneity and flatness of the
present universe requires an extraordinary amount of fine tuning
of the initial conditions in the early universe. One may argue
that since we are dealing with just one unique universe the starting
point could be arbitrarily special. But physicists
dislike fine tuning since it reduces the power of explanation.
In this case let us illustrate just how special the initial conditions
would need to have been.

At the surface of last scatter corresponding to a redshift of
about one thousand we have seen that the density perturbations
are at a level of one in a hundred thousand, already very tiny
but accommodated in inflationary scenarios. Without inflation
these perturbations would originate from very much smaller perturbations
earlier in cosmic time. Roughly we may expect perturbations
to evolve linearly with the expansion of the universe.

Thus if we extrapolate back in time from the last scatter surface
the density perturbations become progressively smaller. Going back
to one second after the would-be Big Bang they have already diminished
by a factor of a million so that they are then only one part
in a hundred billion. This is so close to perfect homogeneity
that it is an extremely fine-tuned situation meaning that it
is almost infinitely unlikely among all initial
conditions. More generally, a theory is considered fine-tuned
if there are extremely small (in this case the relative size of the perturbations)
numbers which are not zero but whose size are not naturally explained.
The technical term "naturalness" was invented for such a case.
Naturalness implies the absence of fine tuning of the parameters
of the theory.

If we go back even further in time, the situation becomes all
the more unnatural. At the Planck time the red-shift has increased
by another twenty-two orders of magnitude so the perturbations of 
the unadorned Big Bang model must then be at the level of a zero
followed by thirty-two zeros and then a non-zero digit! Such
an absurdly small number is taken as a sign that something is seriously
inadequate about the theory. While it does not violate any physical
law to assume such an initial condition it violates 
common-sense and is not what any
self-respecting theorist will include as part of a theory. In
other words, it will accommodate the observations but does not
explain why there is such very high homegeneity in terms of any other
physical principles.

\bigskip
\bigskip

The situation concerning flatness is quite similar. As we
have already discussed the proximity of the present total
energy density to the critical energy density implies
that if we extrapolate back in time the proximity becomes
accurate to the extent that the ratio of the total density
to the critical density becomes exactly one to an
incredible accuracy.

Going back to one second after the Big Bang, this ratio
must equal to one with an accuracy of one in a hundred
trillion, again an intolerable amount of fine-tuning
from the theoretical physicist's viewpoint. At earlier
time it becomes even more intolerable because at the Planck time
the proximity of the ratio to one must have
been precise to better than one part in
a trillion trillion trillion. This seems an inevitable part
of the unadorned Big Bang picture. It does not refute
that general picture, for which there is very much support
including the nature of the cosmic expansion, the
properties of the cosmic microwave background radiation
and the successes of nucleosynthesis of the light
elements.

What is does mean, however, is that an augmentation of
Big Bang theory is needed, particularly for the very
early universe, to replace the assumption of such very
extremely fine-tuned initial conditions.

\bigskip
\bigskip

One can imagine alternatives to inflation which could
explain such fine-tuned initial conditions. One attempt
has been to invoke the collisions of three-dimensional
objects, or "branes", colliding together in a higher-dimensional
space. This can have the effect of inducing the required
amount of homogeneity and flatness just because the
collision effects the whole three-dimensional space
equally. So far such an "ekpyrotic" scenario has run into
a wall of criticisms concerning the technical details
although such an imaginative idea is worth exploring
as a potentially viable alternative to inflation. Of course,
the assumption of extra spatial dimensions is itself a big step
to take! But only time will tell.

Yet another speculative possibility is to assume that the Big Bang
originates from some pre-existing phase which transitions
into the present universe in such a way that the
necessary initial conditions are automatic. Such a "Pre-Big-Bang"
picture could shed light on the deep conceptual problems
of the extremely early universe.

\bigskip
\bigskip

Once we move forward in time from the surface of last scatter
the density perturbations which are
already present grow further and at the same time new ones enter
the horizon. Gravitational instability leads to stronger growth of
perturbations and creation of large scale structures.

Structure is characterized by densities substantially larger
than the mean cosmological density. At present the mean density
is well measured and can be expressed in a variety of units.
The dimension of density is of energy per volume and it
turns out that it can be simply expressed as about ten
electron volts per cubic millimeter. An electron
volt is the energy acquired by an electron in traversing a
potential difference of one volt. This mean energy density
is of course extremely small by any everyday standard. It
is one billion trillion trillion times smaller
than the density of water.

As we look at non-trivial structure the over-density,
which is the technical name for the mean density
of the structure compared to the mean background cosmological density,
increases. For a cluster of galaxies, the overdensity
is about ten to one hundred while for a galaxy it
is ten thousand. When we come down in size to the Solar System,
the overdensity is closer to a hundred billion.
To get an idea of how empty the universe is: if
the Solar System were filled uniformly at the density
of water it would have about the same
mass as the whole universe! 
Bear in mind that the Solar System is infinitesimally small
compared even to the Milky Way and completely negligible 
in size with respect to the entire visible Universe.

If we consider compact objects like the Sun or the Earth, the
overdensity becomes the same huge factor of a million trillion trillion
since their density is comparable to that of water, within a factor of a few.

The conclusion is that, although
there are such very strong fluctuations in density at cosmologically
small scales, if we consider scales large compared to clusters of galaxies,
say more than ten Megaparsecs, then the universe becomes to a good approximation
exceptionally smooth both with regards
to its homogeneity
and it isotropy, being the
same in all directions.

Such large-scale smoothness is all important in concocting an appropriate theory
of time and of the universe
because it implies that the simple Friedmann equation derived from 
the assumptions of smoothness and of general relativity
is very likely to
be the correct choice, and we shall
continue to make this assumption throughout the rest of the book, 

\newpage

\bigskip
\bigskip

\LARGE

\begin{center}

{\bf \underline{ Chapter 3}}   

{\bf STRUCTURE IN THE UNIVERSE}

\end{center}

\normalsize

\bigskip
\bigskip
\bigskip

Having raised the enigma of the smoothness of the universe in
the large, it is now appropriate to address the opposite issue of why 
there exists structure in the form 
of very many galaxies each containing very many stars?
One such galaxy is the Milky Way containing our Sun
around which our Earth orbits once per year. Recent work
provides a plausible
explanation for the origin and evolution of such
structures. The structures are only a small
perturbation of the smooth universe and yet its most fascinating aspect.
The explanation is based on the idea that the structures
are "seeded" in the extremely early universe from effects of quantum
mechanics, a theory formulated to describe the smallest scales
associated with atoms, and now relevant, because
of the enormous expansion of the universe since the Big Bang, to
gigantic structures such as clusters of galaxies.

In the observations of the surface of last scatter 
there is a one in a hundred thousand deviation
from complete smoothness. This tiny effect must lead to the structure formation.
There are two aspects to explain: the primordial origin 
in the very early universe of these fluctuations
and then their subsequent evolution into the observed galaxies and stars.

We first discuss the origin of fluctuations and this 
requires a familiarity with quantum mechanics.
The precursor of quantum mechanics is classical mechanics which was set out
in the 1680s in the Principia, the book written by Isaac Newton.
Based on simple laws such as force equals mass times acceleration,
classical mechanics successfully describes the 
motions of objects resulting from pushes and pulls.
This includes not only objects of everyday sizes
but, when augmented by the law of classical gravity enunciated
in the same book, objects like the 
Moon circulating around the Earth and the Earth
orbiting the Sun. Classical mechanics was so successful 
that Newton's Principia dominated physics for the next two hundred years.
It is impossible now to imagine any single publication 
dominating the physics community for anything like so long.

In 1864, classical mechanics was extended further by
Maxwell's theory of classical electrodynamics which successfully
accounted for the motions of charged
particles in electric and magnetic
fields. The combination of
classical mechanics and classical electrodynamics
formed the foundations of theoretical
physics as it existed at the end of the
19th century. This great edifice of knowledge
was such a source of pride to physicists that at least
one great physicist announced 
around the beginning of the twentieth century
that physics was essentially over
because everything was understood.

\bigskip
\bigskip

Such hubris was short-lived. 
In the early part of the 20th century, two serious limitations
to the applicability of the foundations of theoretical physics
were found, both of which led to revolutionary
advances.

The first limitation was associated with the concept of the aether.
Aether was invented as a hypothetical medium through
which electromagnetic waves could propagate. 
It seemed at the time inconceivable that
such a wave could travel through vacuum and would need aether
as an analog of the air through which sound waves propagate.
But if there is such an aether, the Earth is moving through
it on its orbit around the Sun at a speed of about
one ten-thousandth of the speed of light. Therefore, it
was argued, the speed of light should be different
whether the light travels parallel or anti-parallel
or perpendicular to the Earth's motion. A precise
interferometric experiment was completed in 1887
intended to measure such differences. The surprising result
was that the speed of light did not depend on direction.
The conclusion was that the concept of some medium
such as an aether at rest in the cosmos was erroneous.

This aether conundrum and the negative result
from the interferometer
experiment were in the air, so to speak,
at the beginning of the 20th century and a crucial
ingredient in the invention of special relativity theory.
Relativity leads to the prediction, already mentioned,
that information cannot be transmitted faster than the
speed of light as well as the prediction that the speed of light
is always the same in vacuum for any direction, and for any
source and observer.

The rise and fall of the aether idea may have a current
analog. Many theoretical physicists believe that the
understanding of dark energy, to be discussed in
later chapters, could require another revolution equally
as profound as relativity or quantum mechanics.
The dark energy could play a role in theoretical physics
of the 21st century of sparking a new revolutionary idea
just as the aether cunundrum did in the 20th century.

The second limitation of classical physics 
is the one which led to
the invention of quantum mechanics in the 1920s.
It was first noticed in the properties of 
heat radiated from a very hot body, and was heightened by a lack
of understanding of the stability of atoms.

\bigskip
\bigskip

For example, the hydrogen atom is comprised of a positively
charged proton which acts as the hydrogen nucleus
and a negatively charged electron orbiting around it.
Regarding the much heavier proton as at rest and the electron
circulating around it then, according to classical theory
of electrodynamics, the electron must
radiate continuously and lose energy. After a tiny fraction of a second
it will spiral into the nucleus and the atom will collapse.
The same is true for all
atoms of the periodic table and so classical theory predicts
that every atom in the universe can live for much less than
one second. Needless to say, this total disaster 
for the classical theory was not foreseen
by the great man earlier announcing the end of physics.

This led after some false steps to the invention
of quantum mechanics to replace classical mechanics
especially for describing the inner workings of the atom.
In the limit of large everyday sizes the quantum theory
reverts to the classical theory so that all the previous
successes are maintained. But at the atomic size, one 
important prediction of quantum mechanics is that
energy is not continuous but exists only in discrete
amounts called quanta. Thus the atomic electron does not
radiate energy continuously but in discrete amounts. Once
the electron is in its lowest energy state no radiation
is possible according to quantum mechanics and the stability
of atoms follows. Quantum mechanics succeeds where the 19th century
classical theory fails completely.

Quantum mechanics has many other successes but one particular feature
merits our present attention. According to this theory
there is an uncertainty in the value of all quantities which
would classically be precisely knowable. For example, the
position of the electron in the hydrogen atom cannot be specified
like that of the Moon going around the Earth or of the Earth going around the Sun.
Instead we know only a probability distribution, or wave function,
which tells us what the probability of finding the electron in some region
of space is. The total probability of finding the electron somewhere
is necessarily one just as in the classical case. It is just
that on the atomic scales the position and speed of the electron
have a quantum fuzziness which is a subtle aspect of the theory.

\bigskip
\bigskip

A related phenomenon of quantum mechanics is that the vacuum
is no longer empty but is alive with spontaneous creation
and annihilation of particle-antiparticle pairs which may
live a very short time immediately
to be replaced by others. To describe this situation needs
a marriage of quantum mechanics and special (not general!)
relativity called quantum field theory.
The essential point is that the vacuum value of a quantum
field is continuously fluctuating rather than having a definite
value as it would in a classical description. This
uncertainty is very important for understanding
the structure formation in the early universe. This concept
has no couterpart in everyday life. For an intuitive picture
in the imagination think of a very sharply
focused photograph as the classical theory and then 
put it very slightly out of focus.
To capture the nature of quantum mechanics the out-of-focus
scale needs to be of atomic size since everything is fluctuating
by about that much and not more. 

At this point, there seems to be no possible connection between
quantum mechanical uncertainty and the stars and galaxies. But
that would overlook the enormous amount by which the universe
has expanded.

\bigskip
\bigskip

In the inflationary scenario, as mentioned already, the vacuum
is assumed to roll in a potential between a higher and a lower energy
value as part of a phase transition between an unstable vacuum 
and a stable one.  As the quantum field, technically called a scalar or 
more specifically an inflaton field, rolls down the potential
the universe undergoes a very rapid expansion.
From a purely classical viewpoint, 
at the end of inflation 
the consequent surface of last scatter would have perfect smoothness
properties with no variations at all to seed structure formation.

In quantum field theory, however, the field must fluctuate while
rolling down the potential and these quantum
fluctuations can lead to the tiny ripples on the surface of last scatter.
There are some intermediate steps in realising this idea but it may work
out. To be honest, the exact size of the ripples at one part in
one hundred thousand cannot be predicted from inflation theory.
That is the amount necessary to evolve into the 
structure observed much later as well as being the size of the fluctuations
observed by COBE (1992) and WMAP (2003,2006) measurements of the cosmic microwave
background radiation at the surface of last scatter.
It is perfectly natural the such ripples at such a level 
originate from the quantum mechanical fuzziness during inflation.

The inflationary period is where the scale factor defining the
size of the universe increases by at least twenty-eight
orders of magnitude. If the expansion is exponential in time
it implies that the Hubble parameter is constant during inflation.
The Hubble parameter is the ratio of the time derivative
of the scale factor to the scale factor itself.
A property of an exponential is that its derivative equals to itself.
The comoving visible horizon decreases as the inverse of the scale
factor which means that the fluctuations go outside of the horizon
during inflation. Another way of looking at it is the fluctuations
are expanded exponentially but the Hubble size remains constant.
At a much later time, very long after inflation, the fluctuations 
re-enter the horizon. While outside the horizon the fluctuations
are essentially frozen and re-enter as classical fluctuations
ready to grow linearly with expansion and seed the structure
of galaxies and the stars.
On a clear night when one looks at the
thousands of stars, and the Milky Way meandering across
the sky, there is (only for the last twenty years)
the new insight that they can originate from
quantum fluctuations which happened 13.7 billion
years ago.

These fluctuations from quantum uncertainty during inflation
have three very specific and desirable properties. 
Firstly, they are adiabatic
and seed density fluctuations systematically
in the various components like photons, neutrinos, baryons
and dark matter which are present in the universe.
Adiabatic is a term referrring to the fact that no heat is transferred
into or out of a comoving volume.

Secondly, the perturbations are gaussian which means that the different
wave numbers are all uncorrelated and each has a probabilistic
distribution of a particularly simple and expected type. 
Third and finally, provided the potential
governing the inflaton field during inflation satisfies conditions
technically called slow-roll requirements, the spectral index 
which describes the relative
power in different wave numbers is predicted to be close to one.
All of these three properties are required
to make a successful simulation of observed structure formation.

This is how inflation not only solves the smoothness properties, known as 
the horizon and flatness problems,
but even more impressivly can give a connection between quantum mechanics and the
large scale structure observed in the universe. It is
an appealing idea that the extremes of the very large and the very small
can be so intimately related.

\bigskip
\bigskip

Now we address the second question raised above
concerning structure formation:
how do the ripples on the surface of last scatter grow into the interesting
and beautiful galaxies?

The evolution of structure from the fluctuations
as they enter the horizon involves nothing beyond Newton's law
of gravitation as enunciated in the 1680s.
This gravitational attraction acts principally on the nonbaryonic dark
matter but also on the baryonic component
some six times smaller.

From the time of recombination when the visible universe was one thousand times
smaller than today until when it was merely twenty
times smaller no significant structure was formed. 
This era is picturesquely known to specialists as the dark ages.
After recombination the photon and neutrino components
remain negligible and evolve respectively into the current
cosmic microwave background which is very well
studied and a relic neutrino background which has
yet to be detected.
During the dark ages the fluctuations in all components
begin by growing linearly with the expansion of the universe.

Eventually gravitational attraction causes local perturbations 
of nonbaryonic dark matter to grow nonlinearly
and become more singular until,
according to one scenario, the first baryonic stars can form
with a mass of perhaps one hundred times the solar mass.
Newton's law is adequate for this.
Such massive stars collapse under their own gravitational
attraction.  

\bigskip
\bigskip

As nonlinear effects dominate, the system
must be analyzed semi-analytically or by
computer simulation. Such simulations confirm
that starting from the type of perturbations already discussed,
the end results for the present universe using the
known quantities of nonbaryonic dark matter and baryons
can lead to a structure similar to that observed
in galaxy surveys. The dark matter first
clumps and then the baryons follow suit
to yield clusters and superclusters of galaxies as well as
the large voids which are observed. There are problems of 
detail at small scales; for example, simulations typically give too many
small satellite galaxies for each large galaxy,
and too much dark matter accumulates at galactic cores.
But the three-dimensional distribution of
structure on all larger scales gives rise to pictures
which look to the eye, and 
more importantly to detailed statistical analysis,
indistinguishable from the real universe.

To accommodate structure it is necessary to use
both general relativity and quantum mechanics,
though not together at any particular scale,
with the assumptions of the cosmological
principle and inflation.
General relativity with the cosmological principle gives rise to
the underlying geometry of the expanding universe.
Quantum uncertainties in the inflaton field
provide a possible and plausible primordial origin of fluctuations which exit
the horizon in the very early expanding universe during inflation
and re-enter the horizon much more recently.
The evolution of these fluctuations into galaxies and stars
requires only Newton's law of gravity which is not a separate assumption
but a component of general relativity.

The theories of quantum mechanics and general relativity
are applied separately to different regimes in this picture of theoretical cosmology.
This is fortunate because there is no fully established marriage 
of general relativity and quantum mechanics
into a consistent theory.
The leading candidate is string theory and one very active area
of research is the application of string theory
to theoretical cosmology in a subject called string 
cosmology, a field so new it is still
evolving towards a more precise definition. 
Earlier attempts to connect string theory with particle
phenomenology, which describes the interactions of quarks and leptons deduced at
high-energy colliders, have so far been relatively unsuccessful.
Whether support for string theory will
come first from cosmology
or from phenomenology is an interesting open question. Only time
will tell.

\bigskip
\bigskip

The standard model of particle phenomenology was first invented
in the 1960s and 1970s. First a unification of electrodynamics
and weak interactions was proposed by Sheldon Glashow in 1960 later
completed in 1967 by Abdus Salam and Steven Weinberg in
a form that was conjectured by them to be a consistent
quantum field theory, formally called renormalizable. The fact
that the theory really was renormalizable was proven in the early 1970s
by Gerard 't Hooft and Martinus Veltman. It was shown by them that the theory was
equally as consistent and as amenable to precise unambiguous
calculation as is quantum electrodynamics, the marriage of quantum mechanics with
the classical electrodynamics theory of the 1860s. 

In the 1970s a parallel development was the evolution of a similar field theory
of the strong interactions called quantum chromodynamics or QCD.
The combination of QCD with electroweak theory comprises the standard model. Its detailed
predictions have held up remarkably well. Even fourty
years after the original proposal all experimental data, with one exception,
agree with the predictions, up to an impressive
one in a thousand accuracy. The exception is the non-zero neutrino mass first established
in 1998 and which leads to the necessity of some modification,
still under intensive study, of the standard model.

\bigskip
\bigskip

The standard model led to the successful prediction
of new elementary particles including the charmed quark and, following
the bottom quark discovery, of the top quark. There are six known flavors
of strongly-interacting quark which fall into the three doublets:
(up, down), (charm, strange), (top, bottom). There are also three
corresponding doublets of non-strongly-interacting elementary particles
called leptons: electron, muon and tau with a partner neutrino for each type.
These all group into three quark-lepton "families".

The standard model also led to the prediction of new types of weak interactions
called neutral currents, discovered experimentally at CERN in 1973
and the weak intermediate bosons W and Z discovered, also at CERN, in 1983.
All in all, this theory has been spectacularly successful
and forms the basis of a theory
for at least all the luminous matter seen in stars 
and galaxies and presumably all the baryonic matter. 
At the level of microscopic, subatomic scales it can be said
without hesitation that the standard model
is the greatest achievement of theory
in the second half of the twentieth century.
It is therefore mandatory to ask how extending
this successful model
can accommodate features of observational cosmology
particularly the nonbaryonic dark matter. 

\bigskip
\bigskip

It has been a major industry for at least the last thirty years
to extend the standard model in various directions.
One idea is to unify strong with electroweak interactions 
into a grand-unified theory.
Studies of this type of theory at high temperatures 
actually led to the original idea for cosmological
inflation.
Just as with the uncertainty of how early, or 
equivalently at how high a temperature,
inflation took place, there is an uncertainty in the energy scale
at which grand unification happens. In the earliest and simplest such 
theories the unification scale was very high suggesting an extremely early
inflation merely one trillion-trillion-trillionth of a second after the Big
Bang. There are more recent unification schemes where unification 
as well as inflation could take place at much lower energies 
but, on whether such an alternative is correct, the jury is still out.

One other well-studied extension of the standard model is based on
a serious technical issue within it. There is a scalar particle,
the Higgs boson, necessary for accommodating the symmetry breaking
between the electromagnetic and weak interactions, which gives rise
to violent infinities called quadratic divergences. These seem to render
the theory inconsistent and lead to the consideration of an
extension called supersymmetry.
In such an extension each standard model particle
has a new partner: a quark has a squark,
a lepton has a slepton, and so on. It is not economical but does resolve
the "naturalness" problem associated with quadratic divergences.
It also leads to a candidate particle, known as the neutralino,
which could constitute the nonbaryonic dark matter 
which is established as making up
almost one quarter of the cosmological energy density.

\bigskip
\bigskip
 
The above outline gives just a rudimentary idea of the very strong interrelationship
between particle theory on the one hand and theoretical
cosmology on the other. At an earlier time, say around 1980, 
particle theorists treated cosmology with some condescension because
the cosmological data were so inaccurate compared to 
the reproducible precision data from
high-energy accelerators. This has by now completely changed as the data on the
cosmological microwave background radiation
in 2003 has achieved high-enough accuracy
to be characterized as "precision cosmology". The two
studies of the very small
(particles) and the very large (cosmology) have become inextricably intertwined.
Most university physics departments now combine these two disciplines
in a single group of researchers. The subject of string theory falls neatly
into the rubric of "Particles, Strings and Cosmology",
the title of one well-known series of international conferences.
In academia, strings unify not only the aspects of 
particle phenomenology and theoretical cosmology but also
stimulate common threads of research in physics and mathematics departments.

\newpage

\bigskip
\bigskip

\begin{center}

\LARGE
 
{\bf \underline{Chapter 4}}

{\bf DARK MATTER AND DARK ENERGY}

\normalsize

\end{center}

\bigskip
\bigskip
\bigskip

In this and the next chapter we discuss the present make-up of the universe.
Baryons are the stuff of which everyday
objects are made of so we start with a brief disussion
of that component.

There are two methods to estimate the energy density due
to baryons in the universe.
One which was first used in the 1960s
is by calculating the formation of helium and 
other light elements in the early universe about
one minute after the would-be Big Bang.
The other which has been possible only since 2000 is
by analysis of the relative heights of the odd and even
acoustic peaks
in the anisotropy of the cosmic microwave background
at a time three hundred thousand years after the would-be Big Bang.
These two methods which analyze therefore 
quite different cosmological epochs
agree very well with each other. The result
is that the baryons make up about
four percent of the total critical energy density.
The visible luminous baryons corresponding to the stars
that shine add up to only about one per cent so the other
three percent is invisible and labelled baryonic dark matter.
This four percent of the total energy density
is the only part of which we have a clear understanding.
The present understanding of the "dark" compinents should make
theoretical cosmologists very humble but
what an opportunity for young people entering the field
that we have such limited understanding of ninety-six
per cent of our universe.

\bigskip
\bigskip

Nonbaryonic dark matter comprises about twenty-three percent
of the critical density or some six times the baryonic density.
Nonbaryonic dark matter is much more mysterious than baryonic matter
because it is not something familiar in everyday life. Its presence has
been strongly suspected already since the
observations of Zwicky in 1933 but despite
seventy years of study its nature is still a subject only
of speculation.
Nevertheless, two good candidates for nonbaryonic dark matter
are undiscovered particles hypothesised
by considering extensions of the standard model of
particle physics.

\bigskip
\bigskip

This standard model is a gauge field theory which describes
the interactions of quarks and leptons. It has two pieces:
one describing the strong interactions of quarks and known
as quantum chromodynamics (QCD), the other describing a unification of 
electromagnetic and weak interactions known as electroweak
theory. The two pieces are so far unconnected. Nevertheless, it is
seductive to attempt to unify the strong
and electroweak pieces into a single theory technically
called a grand unified theory. In such a theory all the three couplings
for the strong, electromagnetic and weak forces unify
at a very high energy far above familiar collider energies.

\bigskip
\bigskip

Such a grand unified theory has many attractive features
and leads to remarkable predictions. The most striking prediction
is the instability of the proton which implies that all
objects are eventually unstable. Of course,
the proton lifetime must be exceedingly
long just because material is very stable and even
stronger limits are placed
by the non-observation of proton decay in
dedicated experiments.
Nevertheless there are
certain technical problems with the theory.
For example, the electroweak theory
contains a scalar field known as the Higgs boson which
plays a crucial role in breaking the symmetry between
electromagnetic and weak interactions. This scalar field
must be included in the grand unified theory but then
quantum corrections, technically called quadratic divergences,
naturally force the
Higgs field to have an extremely heavy mass near to
the grand unified scale. But that destroys its capability to play
the appropriate role in the symmetry breaking phenomenon.
Another issue is that phenomenologically the 
unification of the three couplings
does not occur very precisely within the grand
unified theory.

One way of ameliorating both of these two problems 
is to extend the standard model to what is called
the supersymmetric standard model and to
unify it in a supersymmetric grand unification.
The additional assumption of supersymmetry is a mathematically
elegant symmetry though for which there is no experimental
evidence. It has the role of predicting a superpartner
for each particle in the standard model: for
example, for each quark there is a superpartner
called a squark. The quark has spin one half while the
squark has spin zero. Each superpartner has a spin
differing by one half unit from its progenitor particle.

In the limit of exact supersymmetry the particle and
superparticle must have precisely the same mass. But this is
experimentally excluded because, for example, a selectron
at the same mass as the electron would be easy to detect
and is clearly ruled out by experiment at that mass.
Therefore supersymmetry
is broken and the superpartners, if they exist,
must in general be considerably heavier, in general, than their
normal counterpart.

The assumption of supersymmetry has two advantages.
First, when the Higgs boson is incorporated into
a supersymmetric grand unified
theory the Higgs mass is not effected by
quadratic divergences. This is true even with
supersymmetry breaking provided such breaking
is of a special kind known technically as "soft" and
at a scale close to the
scale characterizing electroweak breaking.
Secondly, the unification of the three couplings
becomes phenomenologically 
more precise 
when the superpartners are included
than in the case without supersymmetry.

While admitting that the supersymmetry idea has still
no support from experiment, there is no shortage
of theoretical work along this line and the design
of detectors for the next generation of colliders is arranged
at least partially to find superpartners if they exist.

What has this got to do with nonbaryonic
dark matter in the universe? This needs a little
more explanation to understand how the supersymmetric
standard model seems to provide a good
candidate particle
to play the role of nonbaryonic dark matter.
In the supersymmetric standard model there is a natural discrete
symmetry called technically R symmetry which means
that the number of superparticles is conserved in
every process. Such an R symmetry is necessary in the
supersymmetric grand unification to avoid a proton
decay lifetime too short for agreement with experiment.

One consequence of R symmetry is that there must exist
a lightest supersymmetric particle (LSP) which 
is absolutely stable because there is no lighter
superpartner into which it
can decay while conserving the crucial R parity.
This LSP is typically a linear combination of
superpartners of the Higgs (Higgsino),
the photon (photino), and the B gauge boson (bino). 
Such an LSP is generically called the neutralino.

The neutralino should have a mass of about one hundred
times the proton mass and interact with strength characteristic
of the weak interaction. As such, it provides an example
of a class of particles invented for the purpose
of comprising the nonbaryonic dark matter
called Weakly Interacting Massive Particles (WIMPs).
When one calculates
how many such neutralinos survive annihilation
in the early universe one is
gratified to find that it fits perfectly
the amount of nonbaryonic dark
matter which is observed.
So the neutralino which follows from the assumption
of supersymmetry is a good candidate to be all of the nonbaryonic
dark matter. It works so well that some theorists regard this as a 
main motivation for belief in supersymmetry because the resultant
neutralino gives such a natural candidate for a cosmological WIMP.

\bigskip
\bigskip

There is a second candidate to play the role of the
nonbaryonic dark matter arising from a different extension of
the standard model
of particle phenomenology from a quite disparate line
of argument. It is called the axion and
its motivation is somewhat less than that for the neutralino
but it is worth to describe as yet another example of how
there is such a strong interrelationship
between cosmology and particles.

In the theory of strong interactions, QCD, already mentioned
there is an issue concerning why the strong interactions
respect certain dicrete symmetries. These symmetries are
parity (P) or mirror reflection and CP which is the product
of parity with charge conjugation which interchanges matter
with antimatter. Both of these symmetries, P and CP, are violated
by the weak interaction but appear to be well respected in 
strong interactions.

Nevertheless there is a particular term which is allowable
in the QCD theory which violates both P and CP and of which
the coefficient must be of magnitude
less than one ten-billionth to avoid
conflict with experiments on bound states, particularly
on the electric dipole moments of mercury atoms and neutrons.

One way to solve this problem is to impose and subsequently break
an additional symmetry. This procedure gives rise to an additional
particle called the axion. It can act as the nonbaryonic dark matter.
The axion theory has certain technical
problems such as that the theory appears
inconsistent when combined with gravitational interactions.

The axion mass lies around a trillionth of the proton mass and so is a
hundred trillion time lighter than the neutralino.
These two candidates underline how little we understand the nonbaryonic
dark matter: the range of possible mass for this missing ingredient
actually lies somewhere between the axion mass 
and a million solar masses, a stunning range in mass of 
some sixty-nine orders of magnitude.

Thus, although the neutralino and axion are plausible candidates, 
the true solution chosen by Nature
for the nonbaryonic dark matter remains enigmatic.

\bigskip
\bigskip

Given that almost one quarter of the cosmological energy density
is nonbaryonic dark matter, roughly six times as
much as the baryonic matter, the question is how
to detect it. The arguments involving the study of galactic
rotation curves, of the virial theorem in clusters, and
of the general cosmological observations of the cosmic
microwave background radiation, the large-scale structure
and the type-1A supernovae are all indirect methods. They
imply the fact that there must exist such nonbaryonic dark
matter. But of what is it comprised?

In particular is it made up of weakly interacting massive
elementary particles (WIMPs) with mass of order
a hundred times the proton mass, or the much lighter axion,
or is it made
of far larger entities with the mass of the Sun or larger?
It is a measure of our present ignorance that such extreme possibilities
are still viable. Next, given that this nonbaryonic dark matter
has a specific form, what are its interactions with the baryonic matter?

This issue of the interactions is especially relevant to the
possibility of detection. The key question is how strong
are the interactions with ordinary matter. We know that there is
a gravitational attraction from the method of its indirect detection.
If the WIMP were the neutralino state of supersymmetry
then it must also have a weak interaction with ordinary matter.
In that case one method of detection is to use bolometric detectors,
typically large crytals cooled to very low temperature.
If a WIMP particle strikes the crystal, and interacts,
it may deposit energy in the form of
vibrations, or phonons, in the crystal. Such phonons could
then be detected. Such searches have been made and even
claims of a positive signal have occurred, but
such claims have not yet been reproducible and are therefore
generally disregarded. The detection of the nonbaryonic
dark matter will be an extraordinary claim. Extraordinary
claims need backing by extraordinary evidence and that, so far,
does not exist.

Another possibility for detection is from the process of annihilation
of dark matter with dark antimatter. If we assume this
produces normal photons and electron-positron pairs
then the subsequent positron annihilations could be detected
by their gamma ray emission. Indeed there is a suspiciously
large number of positrons near to the center of the Milky
Way and it has been suggested that these may originate from annilihilation
of dark matter. Unfortunately there are other explanations for
the occurrence of such positrons and so the evidence
for direct detection of dark matter is not yet compelling 
from that source either.

The direct detection of nonbaryonic dark matter could be
rendered extremely difficult (one must never
say impossible!) if its interactions are only gravitational
just because of the weakness of the gravitational interaction.
For example, in such a case even if dark matter does annihilate
against its antimatter, it would produce only more
dark matter in the form of "dark photons" or whatever
dark particles are relevant. It is only an assumption that
the nonbaryonic dark matter posseses significant
non-gravitational interactions with ordinary matter.

In such a case we might have to be content with indirect detection
for some time but that would be frustrating in the sense that
the identification of the constituent particles of
nanbaryonic dark matter would remain elusive and would
not act to motivate the building of models
extending the standard model of particle phenomenology.

A more optimistic scenario is that there are significant
non-gravitational forces and that direct detection will occur
in the near future. This would provide an important
guide to model building especially when the nature
of the interactions between the dark and ordinary
matter are further explored experimentally.

Enthusiasts for supersymmetry are naturally encouraged by
the fact that identification of the WIMP with a neutralino
very naturally gives the correct dark matter density
after the WIMP-annihilations have taken place. Indeed this is taken
by some as a strong suggestion that supersymmetry is correct.
Another such evidence is the improved unification of the
couplings in a supersymmetric grand unified theory.

On the other hand, the main motivation for supersymmetry
is to provide naturalness for the Higgs boson in the sense
of cancelling quadratic divergences in the underlying
field theory. If such naturalness can be obtained without 
supersymmetry then that motivation would be removed, although
supersymmetry still plays an important role
in the construction of superstrings as a potential
theory for quantum gravity. In that case, however, the supersymmetry
can be broken at a very high energy scale, perhaps as high
as the Planck scale, and there would then be no reason to
expect to see sparticles at masses accessible to the next generation
of colliders.

It can be seen from such arguments that the direct detection
of dark matter in non-accelerator experiments could help
clarify precisely what particles to expect to be produced
at high-energy colliders. It is in such a sense that the
interrelationship between cosmology and
particle theory becomes even more important.

\bigskip
\bigskip

We have so far discussed the baryonic component of four percent and the
nonbaryonic dark matter component which is twenty-three percent. Yet
the total energy is consistent with the critical density. This means 
that some seventy-three percent of the total 
mass-energy density has not been listed. 

This additional dominating component has been discovered only since 1998
and is called dark energy. What is the difference of dark energy 
from dark matter?

To understand the distinction it is necessary to introduce
the ideas of pressure and equation of state.
Baryonic matter and dark matter exert zero pressure, the corresponding
particles are essentially at rest and the equation
of state which is defined as the pressure divided
by the density is equal to zero. 
Photons which necessarily travel
at the speed of light do exert a positive pressure
and their equation of state is equal to plus one third.
These two values of zero and plus one third are the
most familiar equations of state applicable to dust and radiation
respectively.
Dark energy is strangely different and, in particular, exerts
a negative pressure. In its simplest manifestation dark
energy corresponds to a cosmological constant with
equation of state equal to minus one, or pressure equal to minus
the density.

Negative pressure has no example in everyday life. Normally,
with positive pressure, exerting a force on the piston confining
a cylinder of gas will compress the gas and increase the positive pressure.
For a cylinder of dark energy, however, the force would
increase the volume, completely contrary to the physical intuition
we develop from everyday experience.

We shall discuss much more about the equation of state for dark energy
in the final two chapters of the book. Here we recall the early
history of the cosmological constant, the precursor
of modern dark energy.

After general relativity was first published by Einstein in 1915,
he followed up with its application to cosmology in his very important 
paper of 1917 which may justifiably be regarded as the start
of what we now regard as theoretical cosmology. At that time, \
it was generally believed that the Milky Way
galaxy was identical to the entire visible universe
and that this universe was a static situation without overall expansion
or contraction. We now know that the universe is really four hundred
thousand times bigger and that the expansion can be seen
only in the much larger theater in which the Milky
Way and all other galaxies are treated as mere points.

Ironically, an observational astronomer named Slipher had
published a few years earlier, in 1912, his result
of observing far more red shifts than blue shifts. It seems
likely that Slipher was seeing the expansion of the universe
but he naturally thought he was looking at stars
within the Milky Way rather than at other galaxies outside
the Milky Way.

\bigskip
\bigskip

In reality, Einstein in 1917 missed a golden opportunity to
predict the cosmological expansion which follow 
quite straightforwardly from his 
general relativity theory combined with the cosmological
principle. Instead, he added a negative cosmological
term which made the theory less elegant but allowed a static
universe. Only much later in 1929 did Einstein realize
from Hubble's publication that the universe is
really expanding. He must have, at least figuratively,
kicked himself for not sticking to the most elegant version
of general relativity.

A moral of this story is that a theorist, including
Einstein, may not take his own theory seriously enough. 
A sequel is that since 1998 the effect known as dark energy
which resembles a cosmological constant,
though now with a
positive sign, has appeared and looks more than likely
to remain as a robust part of cosmological theory.

\bigskip
\bigskip

Should one regard this new development as something
which vindicates the mistake of Einstein
in not leaving the term out from his cosmological
application of general relativity?
This new discovery of an effect looking like a cosmological
constant should not be attributed to the
far-sightedness of Einstein. It arises not from
the expansion of the universe but
from the even more surprising fact that the rate of expansion
is accelerating. Noone had any notion
of this even ten years ago, let
alone in 1917, and it has no
connection to Einstein's motivation for adding
such a term in his theory.

Rather it shows that there is only a small finite number of ways to modify
the theory of general relativity while preserving all of its
symmetry principles. It is surprising that the most economical
version fails but future understanding of the additional
piece associated with dark energy may shed light
on how to discuss gravity correctly and relate it to the other interactions.

\bigskip
\bigskip

One ambitious attempt at a consistent theory
of quantum gravity is string theory and one
place where one would have hoped for insight from string theory
is certainly in the issue of the cosmological term.
This has not been forthcoming and string theory
actually seems slightly to favor a negative rather than the 
observed positive sign for the cosmological constant.
This is still probably not the end of string theory
input, mainly because the theory is still plagued
by the present inability to select between an
astronomical number of candidate lowest-energy states.
It is unclear whether this hurdle to progress
will be jumped over in the near or distant future
in order to confront string theory with the real world.
In any case, even if string theory turns out not to be the
correct theory
of quantum gravity, it has already, by a remarkable technical
argument going by the name of ``correspondence"
between string and gauge theories, provided 
very interesting possibilities for gauge theories that could
address some of the shortcomings of the
standard model of particle phenomenology.

The tiny cosmological constant and the cosmic coincidence
of the dark energy density being now comparable
to the matter density may be explicable in field theory
models or alternatively by exotic ''phantom menace" models with an
unexpected equation of state as well as bizarre 
consequences for the distant future fate
of the universe, as we shall discuss in further detail 
especially in the final chapter.

\newpage

\bigskip
\bigskip

\LARGE

\begin{center}

{\bf \underline{Chapter 5}}

{\bf COMPOSITION OF THE UNIVERSE'S ENERGY} 

\end{center}

\normalsize

\bigskip
\bigskip
\bigskip

The newcomer to the cosmological
mass-energy menu, and the dominant component, is
the dark energy.
Since dark energy is considerably more obscure
even than dark matter, and because dark energy
could play for the 21st century the role played
by the aether for the 20th, the way
in which dark energy
was discovered and quickly established
deserves discussion as it is a fascinating story.

In 1929 Hubble discovered that the recession
velocity of a galaxy is proportional to its distance.
The constant of proportionality, the Hubble parameter,
is now known within a few per cent error to be seventy
in the units mentioned earlier: kilometers per second per
megaparsec.

What had been assumed until 1998 was that the rate of expansion
was decelerating. With no cosmological term, whether the universe
were open, flat or closed the natural expectation from
general relativity is that the rate of expansion
is slowing.

One of the major problems with the evaluation of the Hubble parameter
is the reliable estimation of distance to other galaxies. Until
relatively recently this required a sequence of steps
called the cosmological distance ladder, each step introducing
its own error. Finally the resultant distance and the
corresponding Hubble parameter could be uncertain by a factor two.

This distance ladder can be avoided if there is a so-called
standard candle with known luminosity at a very great
distance, ideally from when the visible universe was, say, half
its present size. Such an object is provided by a supernova
of a particular type, type-1A.

A supernova is a gigantic thermonuclear explosion which
happens when a star collapses under its own gravitational attraction.
The most straightforward case is called core-collapse and does
not involve interaction with any other object.
A particularly consistent type involves instead one member
of a binary pair of stars when one of them becomes a supernova.
This is called technically a supernova of type-1A, because
it has no hydrogen in it spectrum (hence not Type 2) and
because it accretes mass from its binary companion (hence
not Type 1B or 1C).

By the study of nearby type-1A supernovae it is found
that the peak luminosity is simply related to
the rise and fall time in luminosity. This regularity means
that we can estimate the absolute luminosity and
hence distance, by the inverse square law, to such
type-1A supernovae. The red-shift and hence the recession velocity
can be measured also and so very distant type-1A supernovae provide points
on the Hubble plot which are exceptionally distant and 
because they act as standard candles they by-pass
the error-inducing process of the cosmological distance ladder.

The key point is that the supernova type-1A can be discovered
as its light output is increasing then observed as it
reaches a maximum then declines gradually in intensity.
The rise and fall in intensity typically takes from a few weeks
to a few months. Since it is not known where and when
a supernova will occur, it is important to make the discovery
to have a wide-angle telescope which may be quite small
typically four meters in diameter. Once the discovery is
made a very large telescope up to ten meters in diameter
or, in space, the Hubble Space telescope,
can follow the subsequent evolution with greater accuracy.

What has been established phenomenologically for the nearby
supernovae type-1A is that the duration of the light curve
characterizing the increase and decrease of the light
intensity is closely correlated with the absolute value
of the peak intensity: the longer the time period, the brighter
the supernova and vice versa. Thus, observation of the
time dependence of the light intensity translates into
the measurement of the peak intensity and hence the distance.
The theory of supernovae type-1A is not sufficiently developed
to derive this relationship between light curve and peak intensity
from a purely theoretical perspective but empirical data 
strongly support the correlation.

During the 1990s such type-1A supernovae were pursued by
two groups of observers and the data analyzed to give
a consensus on the unexpected result. Instead of slowing down
as everyone expected the expansion rate of the universe
is accelerating. Further analysis suggested that this
acceleration has been happening for
only a fraction of the time since the Big Bang and that the
transition from decelerated to accelerated expansion happened
at some time more recent than when the visible universe was
half it present size.

These data suggest that there is a significant component of
the total energy density which exerts a negative pressure
tending to blow the universe apart. One example is a positive
cosmological constant. 

Fortunately there are independent checks for such a dramatic
and unexpected result. If it were only the supernovae, it could
be that they appear dimmer than expected merely because of obscuration
by intermediate dust clouds. Alternatively there could be a systematic
evolution effect between the very distant and nearby supernovae.
Both of these possibilities already seem unlikely. For example,
the obscuration would be expected to effect different colors
differently and there is no sign of that. Also the
similarity of spectra between the distant and nearby
type-1A supernovae would argue against any
significant evolutionary effect.

The observations of these type-1A supernovae was undertaken beginning around
1992 especially by a group based in Berkeley. 
Presumably, it was not anticipated at that time the extent to which
such study would revolutionize theoretical cosmology. 
In 1995 a second group based in Harvard
was converted to join the chase as an independent group to check on the nature
of the output results. Despite some false
starts the two groups eventually
agreed on the result that the expansion of the universe is
accelerating. This result was completely unexpected and so
it was very important to have the two independent groups
making the observations and analysis.

\bigskip
\bigskip

A second independent approach is to analyze the anisotropy
of the cosmic background radiation. By decomposition into
multipoles one can plot bumps, technically called acoustic
peaks, which reflect the acoustic oscillations
of the baryon-photon plasma immediately
before recombination.

The detailed positions and heights of these acoustic peaks
are sensitive to the values of the energy components and so
can give independent evaluations for the amount of dark energy
as well as of baryons and the dark matter. 
This is because the photon trajectories from the surface of last
scatter to Earth depend on the geometry of the intervening
spacetime, whether it is Euclidean or non-Euclidean
and, if the latter, whether the curvature is positive
or negative. It is possible to find analytical expressions
for the acoustic peak positions in this way. In addition,
there are public software codes which will calculate
the detailed shapes of the peaks for any input cosmological
parameters.

\bigskip
\bigskip

Physically we may say that the causally connected horizon
represents only a small fraction, on the order of
a fourty-thousandth of the full visible horizon.
The angle subtended at Earth by that causally connected
region depends on whether the photons traverse straight lines,
as in a flat geometry, or if their trajectories are bent by a curvature
of space. This will effect which multipole is enhanced
in the observations of the anisotropy. For a flat geometry
the prediction which depends on the time of recombination
is that the multipole equal to about two hundred will
be the location of the first acoustic peak. Further harmonics
will appear at multiples of this value in first approximation.

If the curvature is positive the peak will appear at a lower multipole
while if curvature is negative it will appear at a higher multipole
so this measurement give a direct estimate of the nature of the geometry
back to the surface of last scatter.

\bigskip
\bigskip

The baryon content effects the relative heights
of the odd and even acoustic peaks in a calculable way.
The most detailed such survey was published in 2003 from the
WMAP satellite. It confirms that seventy-three percent of
the total energy is in the form of a dark energy, while
four percent is in baryons and the remaining twenty-three
percent in dark matter. The errors in these percentage 
estimates are quite small.

The WMAP data does not pin down the equation of state
for the dark energy. Although it is consistent with the value
minus one as for a cosmological constant, it allows values
both more positive than this and especially considerably
more negative, down to -4/3.

\bigskip
\bigskip

Finally, there is a third independent means of analysis. From
the large scale galaxy surveys such as the so-called 2dF
Galaxy Redshift Survey and the Sloan Digital Sky Survey
the total amount of dark matter plus baryons can be deduced.
The result is some thirty percent of the critical density
with an error of some three percent. This is in agreement
with the other two estimates. We note that each of these two
galaxy surveys included about 500,000 galaxies.
 
To understand better how the dark matter component is estimated,
it is necessary to examine the gravitational attraction and behavior
within a cluster of galaxies. There is a general theorem,
technically known as the virial theorem, applying to any such gravitationally
bound system, which relates the kinetic energy to the potential energy.
This is used to estimate the amount of gravitational mass in a cluster.
By comparing this with the estimated mass of the luminous matter
one can check for any excess and identify such as dark matter.

At the length scales of cluster and superclusters, just as for
individual galaxies, it is found that dark matter is
present at a level some thirty times
the density of the luminous matter, or very roughly
thirty percent of the critical density with errors
of three per cent.

If we input the observation that the geometry is flat
from the CMB analysis then there is clearly a missing
component, other than the luminous and dark matter,
to be identified with the dark energy.

The important point is that the study of galaxy redshifts and
particularly the virial theorem for clusters of galaxies
points to the necessary presence of dark matter. At larger distance scales
it becomes more important but the consensus is that it does not
approach the critical density even for scales approaching 
the size of the visible universe.
Instead the convergence is on an amount of dark 
plus luminous matter approximately thirty
percent of the critical density.
The precise WMAP data confirms a number twenty-seven per cent
for this quantity.

\bigskip
\bigskip

Taken as a whole there is a cosmic concordance which is difficult
to dispute. All three approaches lead inevitably to the conclusion
that there is almost 3/4 of the universe (seventy-three percent)
in dark matter, twenty-three percent nonbaryonic dark matter
and four percent baryons.

Only the four percent is something we really understand! The
remaining dark side of the universe is truly enigmatic.
Even the baryonic component has a majority which is not
dirctly detectable. At most one per cent of the critical density
can be accounted for by the luminous matter in stars and galaxies.
The remainder of the baryonic component must be in the form
of objects too cold or light to sustain thermonuclear power.
These could be Jupiter-sized objects, brown dwarfs, etc.

Nonbaryonic dark matter could be composed of objects from the
LSP particle, a kind of WIMP, with a mass of about a hundred times
the proton mass all the way up
to objects with a million times the solar mass.
This range of masses which stretches over sixty orders
of magnitude reveals our ignorance on this major component.

Even this level of ignorance pales by the side
of the excitement created by the discovery of dark energy.
This could be merely a reappearance of Einstein's
cosmological constant, albeit with the sign reversed, and
with a totally different motivation from
Einstein's. If so, there is still a fundamental issue
of fine-tuning which remains unresolved.

If the dark energy is something more exotic then its understanding
may well require a revolution in theory. That is why the dark
energy provides so much excitement. The fact that it is so
badly understood provides motivation for pushing again the limits
of human intelligence and creativity.

The end result of the cosmic concordance is that the make-up
of the menu of the total cosmic
energy density is therefore 4\% baryonic matter,
23\% nonbaryonic dark matter and 73\%
dark energy. The errors on these partitionings are quite
small and its now seems well established from the cosmic
concordance that this picture is accurate and
unlikely to change significantly in the future,

The surprising and even slightly embarrassing aspect of this
overall composition, and one may add enigmatic,
is that only the 4\% contribution of
baryons is well understood. The remaining 96\% of the
dark side of the universe is amazingly little
understood. It stands as a wonderful opportunity
for the upcoming generation of physicists to 
seize upon and make their reputation on understanding
dark matter and energy.

The dark matter may be already implicit in certain extensions of the standard model
of particle phenomenology, for example the lightest supersymmetric
partner in the supersymmetric extension of the standard model.
For the dark energy, on the other hand, it may require some further understanding
of the fundamental theory of quantum gravity to
make real progress in its description.

\bigskip
\bigskip

From observations in the near future it will be possible to
pin down the equation of state for dark energy. This means
the ratio of the pressure to the density. The density
is always positive but for
dark energy the pressure is significantly negative
and the question is: how negative?
It is known that the equation of state must be less
than negative one third to accommodate an accelerating cosmic expansion.
It is important to know whether the equation of state
is greater than, equal to, or more negative than negative one.
The future of the universe depends strongly on this equation of state of the dark energy.
Its evaluation will require more accurate data on the
cosmic microwave background radiation and on Type-1A supernovae.

We shall discuss the future fate of the dark energy and thence of
the universe in the final chapter but let us briefly preview
here one of the more fascinating aspects for the case where
the equation of state lies below minus one.
In this case there is the peculiar feature that the universe
ends at a finite future time when the scale factor diverges to
infinity. This has been called in the physics literature
the Big Rip. The dark energy density increases without bound
as a positive power of the scale factor, becoming also
infinite at the time of the Big Rip.

In a philosophical sense, this picture is quite attractive because
it introduces a kind of symmetry between the past and the future
in the sense that both are finite in linear time.

The physics of the final billion years before the Big Rip is
quite interesting even if it is extremely unlikely that, even if
it will occur, anyone will be around to observe it. It has been
computed how the dark energy density increases to such a level
that, assuming that there are non-negligible fluctuations in it,
it will gradually rip apart the existing structure.

Some billion years before the Big Rip, clusters of galaxies are
the first to be erased. Galaxies are similarly dismembered
some sixty million years before the Big Rip. Three months
or so before the end, the Solar System becomes unbound.
Just a half hour before the Rip the Earth will explode.
Finally, a ten-million-trillionth of a second before the
final gasp, all atoms will be torn apart and dissociated.

If this scenario is correct, we are living in a cosmological era
some time between the formation of intricate structure from
an exquisitely-smooth universe and a future time when all
the structure will be torn apart just before a Big Rip.
So it is no surprise that we are living when there
exists complex structure.

\bigskip
\bigskip
 
That is concerning the dark energy properties and we discussed
dark matter in the previous chapter. On both dark aspects out
knowledge is inadequate so perhaps we should look again
at the small four per cent of baryonic matter
which we do understand considerably better, and
there is therefore much more that we can say.

The baryonic matter, like everything else, experiences the gravitational
interaction which is classically well described by the general theory
of relativity. The non-gravitational interactions are well-described
by the standard model of particle phenomenology. 

But even this standard model has plenty of open questions, not
the least of which is the large number of free parameters which
must be matched to experiment. With massless neutrinos there were
already nineteen and now, with the massive neutrino sector, that
number of parameters has increased to twenty-eight. 
Let us go through these parameters
which some all-enveloping theory would be expected to explain.

\bigskip
\bigskip

The masses of the quarks and leptons are twelve quantities which
remain mysterious. Apart from the apparent ordering into three 
quark-lepton families, they might as well be twelve random numbers.
Maybe they are, but physicists would much prefer an underlying theory
to postdict them. That would be a very major step in understanding.

There are mixing angles and phases: three angles and a phase for the
quarks; three angles and three phases for the leptons. These ten bring
the total to twenty-two. The remaining parameters include three gauge couplings:
one for QCD and two for the electroweak theory.
Finally there are the Higgs boson mass, the weak scale (or the mass of the
W boson) and, the most peculiar of all, a parameter called theta bar
which controls an unwanted symmetry violation in the QCD of strong interactions.

Even if all these twenty-eight
parameters could be somehow derived the standard model
still remains inadequately unified because the quite similar theories
of QCD and electroweak forces are disjoint and it is tempting
to put them together in a so-called grand unified theory. One
striking test of this idea is the prediction that the proton,
and hence all matter, is unstable. The lifetime is predicted to
be at least a billion trillion times the age of the universe.
An experiment cannot wait that long, of course, but instead
one can observe the astronomical number of protons in thousands
of tons of purified water for a few years with similar sensitivity.
Such water is usually surrounded by large photomultiplier
tubes and placed deep underground. If a proton decays, for example
into a pion and a positron, the decay particles will travel faster
than the speed of light in water and consequently produce
radiated light to be detected by the tubes.

Such experiments have so far failed to detect proton decay, although
the extraordinarily sensitive water detectors have made revolutionary
discoveries concerning neutrinos. For example, the first extrasolar
neutrinos were so detected in 1987
from a supernova in the Large Magellanic
Cloud lying outside the Milky Way. Similarly, it was first demonstrated
compellingly in 1998 that neutrinos have non-zero mass by studying
the angular distribution of events produced by cosmic rays
in such a giant underground
water detector in Japan originally designed to look for proton decay.

The next step in developing the standard model is likely to be
prompted by the data generated in the Large Hadron Collider at
CERN, scheduled to start producing data in the year 2008. It is expected
that entirely new particles will be discovered including the Higgs boson
and hopefully others. The others could be those predicted by
supersymmetry or large extra dimensions. Perhaps more likely
they will be surprising additional states such as gauge
bosons, fermions and scalars somehow
arranged to cure the unnatural quadratic divergences in the
scalar sector of the standard model.

\newpage

\bigskip
\bigskip

\LARGE

\begin{center}

{\bf \underline{Chapter 6}}

{\bf POSSIBLE FUTURES OF THE UNIVERSE}

\end{center}

\normalsize

\bigskip
\bigskip
\bigskip

It might be hoped that with such a detailed understanding of the history
of the universe we are in a position to predict confidently its
future. That this is not at all the case, we shall discuss in this Chapter.

Such a prediction would be possible were the dark energy component absent
so let us first discuss that hypothetical case. The Friedmann equations
that arise from general relativity under the assumption
of the cosmological principle assign a special role to the critical density.
If the total density of baryons and nonbaryonic dark matter are less than
this then the universe is open with negative curvature and will expand for ever
without coming to rest. The rate of expansion will decelerate.

If the total baryons plus nonbaryonic dark matter density is greater than the
critical density then the universe has
positive curvature and will eventually stop
expanding after a finite time then recontract to a Big Crunch.
Finally, if the baryonic plus dark nonbaryonic densities add
to exactly the critical density then the universe is flat,
the kinetic eenrgy equals the gravitational potential energy and so
it will expand forever
coming to rest at an asymptotic infinite time.
Notice that the division between baryons
and nonbaryonic dark matter
is irrelevant in all
three scenarios since both have the
identical equation of state, namely zero pressure.
In all cases, the rate of expansion will decelerate.

\bigskip
\bigskip

The presence of dark energy spoils such simple prognostications
mainly because the equation of state for dark energy is unknown.
Not only is it unknown but it may depend on time giving
even more possibilities for the future.

The simplest possibility is that the dark energy equation of state
is constant and equal to minus one. This is precisely
the case for a cosmological constant.

For this possibility, while the dark matter density falls like
the inverse cube of the scale factor, the dark energy
density remains constant. Clearly therefore the dark
energy becomes more and more important as time
progresses. At present, by a coincidence, the dark energy and dark matter
have approximately equal roles. In the distant past the dark matter dominated
while in the distant future dark energy will take over.

In this case, the scale factor will grow exponentially with
a constant Hubble parameter and no recontraction can be foreseen. The universe
becomes more and more dilute in its energy density. Thus, although
the dark energy played little role in the formation
of structure in the past, it certainly plays the major role in deciding the
future fate of the universe.

This is the main change in our understanding of the future fate
of the universe in the last five years. It was thought until about five years ago that
the matter content, including just baryons and nonbaryonic dark
matter, was the whole story. Dark energy was unknown.

\bigskip
\bigskip

Let us take a more exotic and theoretically-disfavored case where
the dark energy
equation of state is more negative than minus one and constant.
This is consistent, possibly even favored by the status of the
CMB observations, particularly those by the WMAP collaboration.

This means that the dark energy has an energy density which actually grows
with time, rather than being a constant as in the previous case. It
is not surprising therefore that for this scenario
the dark energy takes over even more rapidly and what happens is that
the scale factor gauging the size of the universe
becomes infinite after only a finite time.

This blow up in the scale factor needs some elaboration. Firstly,
this will not happen any time soon. The earliest 
could be in about ten billion years.
The Earth will have been swallowed as the Sun enters its red giant phase after
another five billion
years so any life surviving on Earth will be have been
annihilated long before this so-called Big Rip.

But if there are observational astronomers safely away from the Sun but still in the 
Milky Way, they will cease to see galaxies outside of our Local Group
because they will have expanded away to infinity. It will be as though
the Local Group
were the entire universe. Ironically, this is similar to the way
the universe was perceived in 1917 when the first real theoretical
cosmology paper was published.

At this time, locally-bound systems held together by
gravitational attraction can continue with their local
time but the notion of cosmological time will end. One could
melodramatically call the phenomenon the end of time.
But it is important to distinguish two kinds of time:
local and cosmological. Only the cosmological version ends
and people locally can still wear watches as usual,
though only temporarily as we shall see.
This ending of cosmology is possible only in the case of the equation
of state below minus one. Why is this theoretically-disfavored?

One assumption typically made in general relativity is that
the energy density is not negative in any inertial
frame related by special relativity. This is called the weak
energy condition and is based
on intuition of what is plausible. With such an assumption,
an equation of state below minus one is impossible.

Nevertheless, it is a question which must be settled by
observation not by theory. At first the observers of the 
type-1A supernovae presented their data assuming the weak energy condition
obtained. But they learned better not to listen to theorists and to
keep an open mind. Plotting the data with an unconstrinaed equation of state
is much more useful to the theorist.
The first data release by the WMAP collaboration studying the CMB
made similar assumptions about the weak energy condition
but they have now re-analyzed their data without such an unnecessary assumption.

\bigskip
\bigskip

If the dark energy is going to require revolutionary new
theory for its understanding then the less prejudiced the input
the better. Clearly at risk, especially in this exotic scenario,
is classical general relativity itself and what could be more
interesting? Of course, such a radical departure is rightly
resisted and regarded as a last resort by conservative
theorists.

But let us pursue this radical possibility even further.
The presence of a negative energy density is suggestive
of instability and that the dark energy may be able
to decay into a lower energy state, for example
one where there is no dark energy. If this is what
is technically called a first-order phase 
transition, like water boiling into steam then
the metastable dark energy is analogous to 
superheated water. Such superheated water is water
carefully heated in a very clean vessel to above
its boiling point. It may remain in this state
for a considerable time but eventually a bubble
of steam will exceed a critical size and expand
to precipitate boiling of all the water.

The idea of a critical radius appears also for the dark
energy but here the critical radius is truly
astronomical, at least of galactic size. This means that
there is a huge barrier to the decay and in this way
one can reconcile the metastability of dark energy
with the fact that it has existed already for billions of years.

\bigskip
\bigskip

As already mentioned, one further aspect
of the super-negative equation of state
is that it leads to an unusual future for
the universe. The scale factor characterizing 
the expansion of the universe
becomes infinite after a finite future time,
maybe just a few tens of billions
of years.

This may not bother planet Earth too much because
after another five billion years or so our Sun will
swell up into a red giant, as its store of thermonuclear
power depletes, and engulf the Earth and any
life surviving here.

Nevertheless this Big Rip expansion would effect small
as well as large objects. Above we mentioned that
people could wear watches as usual. This was not completely
true because, apart from there possibly being no people
to wear watches, the dark energy density grows so rapidly
towards an infinite value that even stars like the Sun,
even if already a red giant, and smaller things like 
watches and individual
atoms may get torn apart by this overwhelmingly
repulsive gravitational force generated by dark energy.
Whether or not this happens depends on the causal structure
of the resultant spacetime which in turn depends
on the fluctuation spectrum of the dark energy.
Present data are consistent with perfect smoothness
of the dark energy but it seems more likely
that there exist fluctuations for it just as there are
for dark matter.

\bigskip
\bigskip

If the equation of state, on the other hand, is more
positive than minus one then the future growth of the dark energy
can be less rapid than for a cosmological constant. 
Especially in a situation where the equation of state
is time dependent such that it actually becomes positive at some
future time. After that time the universe will become
dominated by the baryons and nonbaryonic dark matter
and so will evolve just as in the situation without dark energy.

An equation of state more positive than minus one
occurs naturally in a theory with a scalar field which plays a role
for dark energy similar to that played for
inflation by the scalar inflaton. Such a theory is generically called
quintessence. In some versions it may
be possible that the same scalar field plays the role of the inflaton
and of the quintessence field. This is called quintessential inflation.

There is plenty of freedom in inventing a quintessence
model and it is never completely clear whether quintessence is any more
than a parametrization of ignorance. Nevertheless, it
is a lively topic of research and can give some theoretical frameworks
for comparison with the observational data.

\bigskip
\bigskip

So the fate of the universe is as unknown as
the equation of state of the dark energy. The
most conservative possibility is that 
the latter is constant and
equals minus one. It is then the cosmological constant
introduced by Einstein in 1917 for a completely different reason
and with the opposite sign. 

\bigskip
\bigskip

This still leaves the severe issue of fine tuning in
that the value of the constant is over 
one hundred and twenty orders
of magnitude smaller than would naturally appear in a gravitational
theory.

More exciting surely, and no less likely, is a
radical equation of state more negative than minus one.
Such an outcome of observations would lead to a crisis
in theoretical physics as severe as the one created by
the aether issue over a hundred years ago. It could
then, by the dictum that necessity is the mother of invention,
lead to dramatic and revolutionary progress 
in our understanding of the underlying theory.

It used to be thought, before the discovery of the
dark energy component, that the future fate
of the universe depended simply on whether the matter
content $\Omega_M$ satisfied $\Omega_M >1$ which
gives a positive curvature closed universe which
will stop expansion after a finite time and recontract
to a Big Crunch. Or if $\Omega_M < 1$ there is a negative
curvature open universe which expands forever. Finally
if $\Omega_M = 1$ there is a flat universe which also expands
forever, coming to rest at asymptotically infinite positive time.

The present knowledge is that $\Omega_{total} = 1$ 
(at least very nearly) and
so the universe is flat. This resembles most the $\Omega_M = 1$
flat universe without dark energy. But the dark energy
component introduces a major uncertainty, especially with regard
to its equation of state. 
It is possible that this equation of state will never be known
sufficiently accurately from observation that
one can extrapolate into the future with absolute confidence.
To that extent the future fate of the universe may never be
known with certainty.

One other possible future scenario is that the theory itself will
become more certain. For example, a successful theory of quantum
gravity that explained the past history of the universe
could be trusted to predict the future. This could be from
string theory which is the most promising candidate
for a consistent theory of quantum gravity at this time.
On the other hand there is such a variety of candidate vacua
in string theory, possibly $10^{500}$ or more, that
it may require a better theory to make a definite prediction.

Such a successful theory of quantum gravity would be expected also
to shed light on the origin of the universe. What happened before inflation?
Was inflation inevitable? Was there an eternal repetition
of inflationary eras leading to an infinite
number of universes, called a multiverse?

Some of these questions edge towards the limits of scientific enquiry
and may never (a dangerous word!) be answered with certainty. Some workers use
the principle that our universe must be such as to permit the evolution
of intelligent life. This so-called anthropic principle can
be invoked to account for the values of the dimensionless
physical parameters. The majority of people
find this device unscientific because it replaces physical 
explanation by a principle which involves biology.

Still any theory which is to be trusted to predict the future fate of our
universe must surely be able to provide information pertaining to its
origins. Both questions are valid lines of scientific enquiry. It
can and has been argued that the future fate of the universe
is not within physics because the prediction cannot
be tested. This seems to be a semantic question because in principle
it can be tested if only one has the patience to wait
for a few billion years.

At present we can be sure of our extrapolation back to a temperature
of the equivalent of one hundred
proton masses which obtained some ten-billionth
of a second after the Big Bang. As high-energy colliders
push back the energy frontier this will recede by
another order of magnitude or two towards a trillionth
of a second after the Big Bang. This is still too late
for inflation and evidence for inflation will
necessarily be more indirect. Most compelling would
be observation of the gravity waves emanating
from the inflationary era but for these to be of
detectable strength requires that the charateristic
energy scale of inflation be very high, not too far
below the Planck energy. It remains to be seen whether
Nature chooses such a high inflation energy scale.

\bigskip
\bigskip
 
As we have seen, direct evidence for inflation
is particularly elusive just because, if it occurred, it
did so only in the first billionth of a second,
possibly even in the first
trillion-trillion-trillionth of a second,
after the Big Bang.
Electromagnetic radiation measurements such
as optical or radio telescopes are
sensitive only back
to the surface of last scatter some hundred thousand years
later. Abundances of helium and hydrogen and other light
isotopes can tell us indirectly about one minute after
inflation, still a very long time by early universe standards.
Neutrinos may tell us directly about the same period.

The only possible direct evidence for inflation would
seem to be possible by detecting weak gravity waves emanating from
that era especially if the inflation takes place extremely early
on. This is an exciting possibility for which we will have to wait
for at least a few more years.

\bigskip
\bigskip

The discussions of the fate of the universe can be
criticized as untestable because it will be many billions of years
before the actual event. It can even be said that the
discussion of such matters is not even physics, or science.

This seems unnecessarily semantic since if a theory
is successful in accommodating what is known about the past
it is very natural to ask what it predicts for the very distant
future. The time scales involved in cosmology are typically
gigantic compared to the human lifetime. But in physics,
especially cosmology, there is nothing significant about
the biological scale of a human lifetime. It is equally
valid to ask what happens to the universe in the next
hundred billion years as to enquire about the decay
of an unstable particle in a tiny fraction of a second.
There is no difference in principle.

The separation of physics from biology is called into
question also in the use of the ``Anthropic Principle":
that the universe must be such as to allow intelligent
life which can observe it. The use of this principle is
quite controversial since it is a way of avoiding a physical
explanation for certain aspects and parameters describing
our universe.

While it is true that without intelligent observers the science
of cosmology would be impossible, it seems to be a ``cop out"
on scientific explanation to appeal to such an idea to
say, for example, that the lifetime of the universe
must be above some minimal value or that the couplings
and parameters in the underlying fundamental theory
must be within small ranges around their observed
values or else the cosmic evolution would be so different
as to disallow the creation of life.

So it is the preferable path to study physics without the
need to input facts from biology with the idea that
the occurrence of intelligent life is not of central
importance to the physics rather than constraining the
physics such that life is possible. This attitude is
more likely to lead to explanations which are satisfying.

As we have seen, in the minimsl Big Bang scenario, the previous history
of the universe is finite with an age of
13.7 billion years. In some future
scenarios, particularly where the dark energy has
an equation of state less than minus one, the
cosmos ends at a finite time in the future
when the scale factor diverges to infinity.
This provides a philosophically pleasing symmetry
between the future and the past.

One possibility, for which there exists no
substantial evidence but which has
some theoretical appeal, is that the universe
is finite also spatially. This can occur if the
universe has non-trivial topology in space.
If this were the case, it would provide even a 
more symmetric view between space and time.

\bigskip
\bigskip

The recent growth in knowledge about the universe has been
astonishing but, as usually happens, answering several questions
gives rise to others. At present there are
great enigmas about the universe. Perhaps the most striking
mysteries are: what is the dark matter? 
what is the dark energy?

The existence of such enigmas is very healthy for research
in the field because they signal the distinct possibility
that revolutionary advances in the theory, equally
as dramatic as relativity and quantum mechanics were in
the previous century,
will be necessary. When old ideas are tried and fail
it is all the more likely that intelligence and creativity
of theorists is the only avenue to advance knowledge
and understanding.

\newpage

\bigskip
\bigskip

\LARGE

\begin{center}

{\bf \underline{Chapter 7}}

{\bf ADVANTAGES OF CYCLIC COSMOLOGY}

\end{center}

\normalsize

\bigskip
\bigskip
\bigskip

We have, by now, discussed how the past and the future lifetimes
of the universe may be finite by virtue of the Big Bang
and the Big Rip respectively. The Big Bang was first discussed
in the 1920's and some physicists found it disturbing
that the density and temperature of the universe apparently
become infinite at a finite time (now known as 13.6 billion years)
in the past.

This concern led several of the leading scientists to explore
an alternative where the universe 
cycles between expansion to a turnaround then contracting to a bounce,
such that the temperature and density always remain finite.
This happens an infinite number of times thus eliminating
any start or end of time.

However, the theorists in the 1920s and 1930s could not construct
a model with this cyclic property. The principal obstruction
was entropy and the second law of thermodynamics. During each cycle
the entropy necessarily increases so that subsequent cycles
become longer and expand to a larger size. Extrapolating back
in time the cycles become shorter and smaller until eventually
one arrives a Big Bang again with the same issues of initial
conditions as in the non-cyclic universe. This property removes
the motivation to study the cyclic scenario.

As we have discussed at length in the two previous chapters, the
accelerated expansion of the universe discovered only in 1998
requires a large fraction, over seventy percent, of the Universe
to be in the form of dark energy. This is something we now know
which the theory giants of old (Friedmann, LeMaitre, De Sitter,
Einstein, Tolman) did not know. Therefore, we may consider whether
dark energy can help overcome the obstruction which entropy
provided to oppose construction of a consistent cyclic
cosmology which could preclude a start and an end.

To understand how this problem is addresses, we need to explain the
concept of entropy. Consider a room full of air molecules. The number
of molecules is extermely large, typically of order $10^{30}$ or
one million trillion trillion. Written out this is
$1,000,000,000,000,000,000,000,000,000,000$ molecules! These molecules are
moving with high velocity, typically about 300 meters per second
and constantly colliding with one another.

One might think that describing such a system is hopeless, which
it is in detail such as the motion of any individual molecule.But the
very large numbers lead to simple laws for the entire system.
This is the physics of thermodynamics and staistical mechanics
of which a principal architect was Boltzmann.

\bigskip
\bigskip

Let us think of the following question. The air molecules are
about twenty per cent oxygen and eighty per cent nitrogen. The
oxygen is crucial to us to breathe and survive: without any we
will pass out within a few minutes, and then die. How do we
know that the volume surrounding our nose and mouth will not
be depleted of all oxygen molecules for a few minutes? Why
are there not ambulances waiting for just such an emergency?
Anwering this question will introduce the concepts
of entropy and the second law of thermodynamcs which play
such a central role in cyclic cosmology. 

The molecules is a room can be in an astronomical number
of configurations with regard to the positions and velocities
of all the individual molecules. Statistical mecahnics is the
study of such configurations and its basic assumption is that
every configuration is equally likely. The system will quicky
evolve into a configuration which corresponds to the 
maimum probability. Such configurations are those of highest
entropy which is a measure of he disorder of the system. The
highest entropy states are favored by an enormous factor.

For such an equilibrium state of highest entropy one
can calculate the velocity distribution of the molecules
and the temperature to extreme precision because of the
very high statistics involved. Coming back to our question
the configuration with no oxygen molecules in the vicinity
of a person's nose and mouth for a few minutes is a highly
ordered state of lower entropy and of extremely low probability.
Boltzmann's second law of thermodynamics (the first and the
third are not relevant here) states that the entropy of
an isolated system always remains constant or increases;
entropy never decreases. Thus a room full of air molecules
will never lower the entropy to this unlikely configuration.

We can calculate the probability of it happening using
statistical mechanics. The result is the following, as I
have sometimes asked to a class: if I write the probability
as $0.0000000......$ on the blackboard with each zero
occupying one inch where will be the first non-zero
digit? Will it be at the end of the blackboard? At
the other side of campus? a hundred miles away? The
amazing answer is that it will not be the visible universe so
no ambulances are necessary.

Transition to such a configuration would violate the second
law of thermodynamics because the entropy would decrease.
It is important, however, to emphasize that such an unlikely
configuration would not violate any fundamental law of
physics in the microscopic regime. The second law is only
a statistical law, yet because of the huge number of
molecules the statistical law might as well be an exact
one for practical purposes. Historically it was this subtle
distinction bewtween an exact law which is never violated
and a statistical law which is {\it practically never}
violated for a system of a million trillion trillion
molecules that led to only slow acceptance of Boltzmann's
profound result, especially by mathematicians,
possibly contributing to his eventual
suicide.

Another central aspect of statistical mechanics is provided by
phase transitions such as from ice to water or water to steam.
The thermodynamics of phase transitions was first systematized
in a general treatment of staistical mechanics by the earliest
american professor of theoretical physics, Gibbs. 

\bigskip
\bigskip

The ideas of entropy and the second law can be applied to the universe
as a whole. The entropy of the present universe can be estimated as
$\sim 10^{88}$ which is a one followed by eight-eight zeros. This entropy
lies mainly in the radiation, also in the matter both dark and luminous.
As the universe expands the entropy necessarily increases according to
the second law. The entropy of the radiation component remains constant
and we say the it expands adiabatically; the entropy associated with
matter graudally increases as a result of irreversible processes.

In the very early universe,  entropy increased during inflation
by  a factor $\sim 10^{84}$, so comparison with the present entropy
reveals that the entropy at the beginning of inflation was extremely
low, $\sim 10^4$. This last number is essentially zero on the scale
of the later entropies. In our discussion we will come across two
magnitudes of the entropy: $ \geq 10^{88}$ will be called {\it large}
entropy, $ \leq 10^4$ will be called {\it small} or
{\it essentially zero} entropy. As can be seen, a cyclic universe
with periodic entropy must involve a dramtic {\it decrease}
in entropy to compensate for the huge increase in entropy
at inflation, but how can any decrease be consistent with the
second law of thermodynamics which demands increasing entropy?
 
This is the question that stymied progress toward an oscillating
universe in the 1920s and 1930s. But why did the cyclic universe
seem an attractive alternative to the non-cyclic scenario?

Once it was realized by Friedman in 1922 and independently
by LeMaitre in 1927 that the equations of general relativity
led naturally to an expanding universe, not a quai-static
one as originally proposed by Einstein in 1917, it was also
seen that there was one undesirable feature. Namely, as
one extrapolates into the past at a finite time (now
known as 13.7 billion years) the temperature and density become
infinite, and the scale factor reduces to zero.
This leads to the idea of the explosion of
a primordial "atom" or a Big Bang at the initial time $t=0$.
But the problem is
that the classical equations cease to be applicable 
at this singularity, and general relativity cannot hold.

One reponse is that at a sufficiently early time, the
Planck time, $\sim 10^{-44}$ s, after the Big Bang, the
effects of quantum mechanics must enter so that, in any
case, the classical Friedman equations fail and 
singular behavior at $t=0$ with infinite density
and temperature  and vanishing scale factor could 
be avoided if we knew a satisfactory theory
of quantim gravity. Such a complete theory is unknown
but in attempts at such quantum cosmology various
attempts have been to formulate satisfactory
initial conditions for the quantum version of the
Big Bang. Without entering into techicalities, the
most studied such ideas are due to Vilenkin, and
to Hartle and Hawking. It is fair to say that the
jury is still out on these valiant attempts
to circumvent the Big Bang singularity.

At the classical level the Friedman equations
make a hypothesis known as the Cosmological Principle.
This has two components: (i) the universe is assumed
perfectly homogeneous, and (ii) it is assumed to
be perfectly isotropic or rotationally symmetric.
Now it is clear that neither component of the
Cosmological Principle is exactly valid at smaller
scales such as the size of galaxies and clusters of galaxies
but at extremely large scales, an order of magnitude
above cluster sizes, it does seem to be a good approximation.

Nevertheless, a legitimate question for classical
general relativity is whether relaxing either homogeneity
or isotropy or both could avoid the Big Bang singularity?.
An answer to this queation was offered in the 1960s
by Hawking and Penrose who showed under general
conditions that a past singularity was inevitable
and did not depend on the Cosmological Principle.

However, the Hawking and Penrose no-go theorem
necessarily made assumptions, one of which will be seen
to be important. They assumed the physically plausible
requirement that the energy density never be negative,
since such an energy density seemed to have
no physical interpretation. As we shall see, it is
this assumption which must be avoided in making
a workable cyclic cosmology.

\bigskip
\bigskip

Going back to the 1920s, it seemed desirable
to avoid the Big Bang singularity even classically.
One attractive possibility was a cyclic theory in which the
density and temperature remain always finite and the scale
factor is never zero. 
We have already discussed the independent attempts by 
Friedman and LeMaitre respectively in 1922 and 1927. 

It is significant that
when he heard a seminar at California Institute
of Technology in 1931 by LeMaitre on his research
Einstein was very enthusiastic in his reception
of the cyclic ides.
De Sitter and Einstein published jointly on the topic in 1932.

A particularly clear and endearingly modest discussion of the
role of entropy in cyclic cosmology
is in the book
{\it Relativity, Thermodynamics and Cosmology} by
Richard Chase Tolman. 

To add a personal anecdote, when I was studying for the Final
Honour School in Oxford in 1965 and had immediate access
to several hundreds of physics books, my personal favorite
book was always Tolman's. 
I do recall spending hours then intrigued by the apparent
contradiction between the attractive idea of cyclic
cosmology and the second law of thermodynamics; the
contents of Tolman's book, however, did not appear
on my examination!

\bigskip
\bigskip

One important fact about the universe, discovered only in 1998,
and obviously unknown in the 1920s and 1930s is that the
expansion is actually accelerating. As we have previously
discussed, this led to the identification of over seventy
per cent of the energy as dark energy. This leads to
out question of whether dark energy can intercede
in the apparent contradiction?

Before explaining a positive response to this question, let
us make some general considerations of cyclicity
and entropy in an attempt to make the solution seem
inevitable.

A cyclic universe goes through four stages:
Bounce $\longrightarrow$ Expansion $\longrightarrow$
Turnaround $\longrightarrow$ Contraction $\longrightarrow$
Bounce, repeated an infinite number
of times. 

During expansion, entropy initially
increases by an enormous factor $\sim 10^{84}$
during inflation then more gradually increases
due to irreversible processes in the matter
component. We recall that the radiation expands
with constant entropy and the dark energy
has zero entropy.

Before discussing the turnaround, consider the contraction
phase which presnts its own peculiar issues. One
is that matter and dust will form structure more
readily than during expansion. Black holes, if present,
will expand and more will form, further impeding any
smooth contraction.

\bigskip
\bigskip

There is an even more serious problem with matter content during
contraction, namely that several phase transitions must take
place {\it in reverse} to proceed successfully back to a bounce
in the early universe. For example the phenomenon of recombination
would require, in reverse, that neutral atoms dissociate into
protons and electrons as the temperature increase. This would
decrease entropy and be statistically impossible as a violation
of the second law of thermodynamics.

\bigskip

There are other phase transitions such as the so-called quantum
chromodynamics transtion at a few hundred MeV where quarks and
gluons become confined into hadrons. Also, the weak transition
at about one hundred thousand MeV at which electroweak forces
separate into weak and electromagnetic forces.

How can a contracting universe pass in reverse through such phase
transitions without violating the second law of thermodynamics?

One possibility which is seductive but which we shall quickly
discard is that during contraction the "arrow of time"
reverses. This is an assumption made to argue that entropy
may decrease during the contracting phase.

What does it mean? The arrow of time refers to the second law
of thermodynamics of a staistical mechanical system 
for which increase of entropy defines a "forward" direction
of time. At a biological level, we remember the past but not the future,
and we grow older so for us there is surely a well-defined arrow
of time. For statitistical systems in physics the arrow
of time is equally well defined by the behavior of the entropy. 

The entropy problem could be solved very simply if we were allowed to
adopt a reversal of the arrow of time during contraction and entropy
could then decrease. If we consider this further,
however, it is only a semantic device corresponding
to unacceptable physics although it has been
used by some authors (including, for a few days, by this one!) as a 
last resort to address the thorny cyclicity problem. In a universe
with a reversed arrow of time, statistical mechanics would
become nonsensical because equilibrium states of, say, an
ideal gas would proceed to diseqilibriate into very unlikely
configurations, evidently nonsensical. 

Reversal of
the arrow of time is therefore an absurdity better to avoid if
at all possible. Fortunately we shall find a solution to
cyclicity in which the arrow
of time keeps moving forward in the conventional sense.

\bigskip
\bigskip

To proceed toward an acceptable solution, we need to introduce
a couple of new concepts: branes and causal patches.

\bigskip

Branes were first emphasize in the 1990s on the basis of
sting theory, particularly by Polchinski. String theory
was believed in the 1980s to be a theory of one-dimensional
extended objects, namely strings, interacting with one another.
As such, it was already a remarkably consistent geenralization
of quantum field theory which included within it both gauge
field theory and general relativity. Moreover, it provided
a finite theory of quantum gravity for the first time.

Around 1990, however, it was shown that a theory of only
one-dimensional strings is not internally consistent
and must include higher dimensional "branes", a shortened version
of the word "membranes". In this language, strings are
1-branes. In a ten dimensional supersting theory branes
appear in addition as p-branes for all p between 2 and 9. Which values
of p appear depends on the particular superstring considered;
for example, a 9-brane in 10 spacetimes dimensions is a space-filling
brane.

Of special interest to cosmology is the 3-brane. This notion
has by now been abstracted from string theory leading to the idea
of a brane world in which non-gravitational physics is restricted
to the 3-brane on which we are assumed to live. Strong
and electroweak interactions are restricted to this "TeV"
brane, while gravity alone spreads also into the extra
spatial dimension or dimensions in which the TeV 3-brane
happens to be embedded. In the theory
of brane worlds one remains agnostic about the number of extra
space dimensions which can vary from one up to the six
predicted by superstrings or seven as allowed in M theory.
In our discussion we shall for simplicity assume only
one extra space dimension, or a total of five spacetime
dimensions.

One popular version of the brane world involves two parallel
3-branes where one is the TeV 3-brane and the second is
a "Planck" 3-brane from which gravity originates. This
theory offers an attractive expalnation of why gravitation
is so weak in our observed world. It is remarkable,
for example, that the electric force between two protons
separated by, say, an atomic size (or any other separation)
is some fourty orders of magnitude larger than the
gravitational attraction; that is a one followed
by fourty zeros. Another reflection of the weakness
of gravity is that a small magnet will hold an object
to a refrigerator door against the gravitational pull
of the entire Earth.

For cosmology, our universe will be the TeV brane so in such a brane
world one can derive a Friedman equation for this
TeV brane which has a crucial modification.
Without going into technical details, a new term
appears which involves a parameter with the dimension
of a density and this critical density signals when
the expansion of the universe will stop and turnaround
inot a contractin mode.
This same critical density determines what will be the bounce 
temperature at which the contracting universe will stop
and bounce into a new expanding mode.

The critical density is related to the brane tension and
the new brane term in the Friedman equation plays a
central role in our solution for the cyclic universe.

\bigskip
\bigskip

A central role is played equally by the concept of causal
patch.

\bigskip

To introduce the causal patch we must revisit the Big Rip
where the present expansion of the universe ends at a finite time
in the future when the universe rips itself apart and
the dark energy density, as well as the scale factor
diverges to infinity.

As one approaches the Big Rip, the structures in the universe
successively disintegrate beginning with the largest scales.
An approximate rule is that once the dark energy density reaches
the average density of a gravitationally bound system
that system will become unbound. This implies that first the
clusters of galaxies disintegrate followed by the galaxies
then the solar system. Nothing is immune to this disintegration
process and eventually even atoms, atomic nuclei and
nucleons - protons and neutrons - will become unbound.
These are just the smallest systems we currently know and if there
exist yet smaller higher density bound sytems they too would
be subject unbinding process. 

As another example which is important for the sequel, black
holes are themselves torn apart on the approach to the Big Rip.

\bigskip
\bigskip

At a time somewhat later than the unbinding of a  system, the
bound components become causally disconnected, meaning that
they cannot communicate even at the speed of light before
the universe ends. Eventually we may regard the universe
itself as disintegrating into a huge number $> 10^{84}$
of causal patches which are disjoint and separate. The
idea now is to delay the brane induced turnaround until
a trillion trillionth of a second or less before the
would be rip. At this time, one causal patch contains no
quarks or leptons and certainly no black holes. 

The causal patch at the turnaround contains only dark
energy and a small number of low energy photons. Its entropy
is {\it small} in the sense discussed earlier, meaning
it may have entropy equal to ten, or some such number.
Before the unbinding process, on the other hand, the
entropy was {\it large} by the same discussion, meaning
it was at least a one followed by eight-eight zeros.

This is essentially the reverse of the vast increase
of entropy experienced during inflation and so
may naturally be called deflation.

\bigskip
\bigskip

We have now assembled all the pieces of our proposed
cyclic universe. It will involve ending the present expansion
at a finite time in the future, typically
of the order of a trillion years if the equation
of state of the dark energy is just below minus one
say, $-1.01$.

At the turnaround, only one causal patch will be retained
as our universe and it contracts with constant {\it small}
entropy: the radiation or photons contracts with no
increase of entropy and the dark energy has zero entropy.

What is equally important is that the contracting universe
is empty of all matter whether dust or black holes and
thus can contract without confronting all the difficulties
discussed earlier such as enhanced structure formation,
and growth of black holes as well as formation of new black holes.

The contracting universe, much smaller than its
expanding
progenotor, continues until the
dark energy again reaches the critical value . Then 
the contraction ceases at a certain bounce temperature
and the universe cycles into inflationary expansion
again with the entropy increasing from
a {\it small} to a {\it large} value. The temperature
at which this happens is related to the critical density
which is a free parameter in the model.

\bigskip
\bigskip

This provides a possible model for a cyclic universe which respects
the second law of thermodynamics.

It has two ingredients which could not have been foreseen in
the 1920s and 1930s by leading theorists who attempted such
a model. The first is dark energy which was deduced from
the observed accelerated expansion rate first in 1998.
The second is the idea of a higher dimensional brane world
which has emerged in theoretical physics equally recently.

\bigskip

Another hurdle of more recent vintage was the singularity
theorems of Hawking and Penrose from the 1960s. These assumed
an everywhere positive energy density while a dark energy with
equation of state less than minus one violates such
an assumption.

\bigskip

Only time will tell whether the cyclic model we have discussed
will survive closer scrutiny. A general observation, however,
is that a cyclic universe which avoids classically any
singular or infinite behavior in density or temperature, and which
avoids any beginning or end of time, seems
more aesthetically acceptable than a cosmology which
originates from a singularity.

\bigskip

This new viewpoint on the future of the universe
is possible only given the amazing developments in
both theory and observations since the
very end of the twentieth century.
The conclusion is that conventional cosmology
with time starting at the Big Bang and continuing
for ever in an infinite expansion
now has plausible alternatives, especially this cyclic
model where time never begins or ends.

\newpage

\bigskip
\bigskip

\LARGE

\begin{center}

{\bf \underline{Chapter 8}}

{\bf SUMMARY OF ANSWERS TO THE QUESTIONS:}

{\bf DID TIME BEGIN? WILL TIME END?}

\end{center}

\normalsize

\bigskip
\bigskip
\bigskip

There are three different futures for the universe which we have discussed.
One key question is whether the present expansion phase will continue for
an infinite time, as in the conventional wisdom, or will it stop after
a finite time of order a trillion years? If it stops, a second question
is whether our universe will then contract and bounce cyclically?

\bigskip

Although we cannot say with certainty which future is correct, progress in
addressing this question has been so rapid that it is possible to order
these three futures, according only to aesthetics, in {\it decreasing}
probability with conventional wisdom third. 

\bigskip

The scenarios in {\bf 1.} and {\bf 2.} assume the equation of state of
dark energy $\omega$ is slightly below $-1$, say $\omega = -1.01$, while
scenario {\bf 3.} adopts $\omega=-1$ exactly.  

\bigskip

\noindent {\bf 1. Most likely.} The present expansion will end after a finite time,
the universe will contract, bounce and repeat the cycle. A cyclic universe. 
\underline{Time had no beginning and will have no end.}
This presumes that the entropy problem
has been resolved as discussed in this book.

\bigskip

\noindent {\bf 2. Next most likely.} The present expansion will end after a finite time
in a Big Rip. 
\underline{Time began} in the Big Bang some 13.7 billion years ago 
\underline{and will end} some trillion years in the future.

\bigskip 

\noindent {\bf 3. Least likely.} The present expansion will continue for an infinite time
as for a cosmological constant. 
\underline{Time began} 13.7 billion years ago \underline{and will never end.}
This is the prevailing conventional wisdom which is here being challenged.

\newpage

\bigskip
\bigskip

\begin{center}

\bigskip
\bigskip

\LARGE

{\bf GLOSSARY}

\end{center}

\normalsize

\vspace{4cm}

{\bf Aether}

Hypothesized, but later discredited, medium through which
{\bf electromagnetic radiation} propagates. In physically interpreting
Maxwell's equations it was thought that for light to propagate
through vacuum, as from the Sun to the Earth, it was necessary for
there to be ``something'', the aether, through which the waves
were transmitted. One consequence was that light should travel at
different speeds according to whether the direction was parallel,
anti-parallel, or orthogonal to the motion of the Earth through
the putative aether. A landmark experiment in Cleveland, at what
is now Case University, by Michelson and Morley in 1887
showed that the speed of light did not so vary with direction,
thus strongly disfavoring the aether hypothesis.

{\bf Baryonic matter}

Matter, including the stuff we and all everyday
things are made of, comprised by mass almost entirely
of baryons: protons and neutrons. This forms about 4\%
of the energy of the universe, and is the only major
component which is well understood. The microscopic physics
of this component is well described, at least up to
energies of hundred times the proton mass, by the
standard model of particle phenomenology.

{\bf Big Bang}

By using the Friedmann equation and the known values for the
contributions of matter and radiation on the right-hand-side
thereof, the expansion history of the universe can
be reliably traced back to
one ten billionth of a second after the would-be Big Bang
when a primordial fireball is often regarded as exploding.
In cyclic models this Big Bang is avoided but
it is Widely regarded, though not in this book, as the beginning of 
time some 13.7 billion years ago.

{\bf Bounce}

The transition from contraction to expansion
phases in a cyclic cosmology.

{\bf Branes}

Multi-dimensional objects which occur in {\bf string theory}.
These objects were identified in string theory
in the 1990s and have led to ideas 
about the weakness of the force of gravity
especially that gravity alone propagates in extra dimension(s)
while the other forces are confined to our four-dimensional
brane which bounds the higher-dimensional ``bulk'' space.
In particular, attempts have been made to replace inflation,
couched in the language of quantum field theory in
four dimensional spacetime, with a higher-dimensional
brane-world set-up where the Big Bang is initiated by
the collision of two branes in an extra space dimension.

{\bf CERN}

European laboratory for high-energy physics near Geneva, Switzerland.
Under construction there is the largest particle collider
in history in which protons will collide with total energy
of 14,000 times the proton mass. It is expected
that the elusive Higgs boson will be discovered,
together with (unknown) other particles which will
hopefully shed light on the mysteries of the standard model. 
As well as its large complex of particle accelerators, CERN
has a theory group which, if one counts both CERN employees and visitors,
is the largest group of particle theorists anywhere in Europe, probably
the world.

{\bf COBE}

COsmic Background Explorer. This was a satellite experiment
to measure the cosmic microwave background (CMB)
and was the first to detect the small
anistropy at a level of about one in one hundred thousand.
These small perturbations seed the formation of 
large scale structure
in the universe including galaxies and stars. 

{\bf Cosmic microwave background}

Gas of photons which fills the universe.
The study of this radiation is one of the most fruitful
methods to investigate the early universe
from the time of some 400,000 years after the Big Bang. 
At this {\bf recombination} era the universe became transparent to photons
and they propagated unimpeded. The gas
is thoroughly
eqilibriated and has a spectrum corresponding now to
a temperature of 2.725 degrees above absolute zero.
The CMB was originally discovered by accident
in 1965 by Penzias and Wilson who were looking
for anisotropic microwaves correlated with the plane of the Milky Way.
 
{\bf Cosmological principle}

The assumption that the universe is 
{\bf Homogeneous} and {\bf Isotropic}.
On the largest scales, these properties are
well approximated by the observed universe and
using this cosmological principle simplifies
the differential equation, the Friedmann equation,
characterizing the expansion of the universe.

{\bf Cosmology}

The scientific study of the universe. This includes the
finite time back to the Big Bang and the finite or
infinite time future discussed in this book.
The beginning of modern theoretical
cosmology can be taken as the paper, despite its flaws,
by Einstein in 1917 where he first applied general relativity
to the whole universe. One cannot help wishing he had
anticipated the work by Friedmann, Lema\^{i}tre and
Hubble in the 1920's: what a paper that would have been!

{\bf Critical density}

A special value of mean energy density of the universe.
Since the WMAP data was released in 2003 we now believe the
actual energy density is very close (within 2\%), and 
possibly extremely close, to this value, meaning
that the geometry of the universe is flat or
Euclidean. If the density is larger (smaller)
than the critical value the universe is closed (open)
and has positive (negative) curvature. Such closed and open universes
were widely considered until the recent observational
discovery that the universe is flat.

{\bf Cyclic Cosmology}

To avoid an infinite density and temperature at the Big Bang
theorists since the 1920s have studied the possibility of
cyclic cosmology where these quantities remain finite
but the universe expands to a turnaround then contracts to
a bounce and again expands. If this cycle can repeat an infinite
number of times it avoids any beginning or end of time. For
decades an apparently insuperable theoretical hurdle
was the second law of thermodynamics but the discovery
of dark energy has led to renewed interest in the idea. 

{\bf Dark matter}

About 27\% of the total energy of the Universe. It is comprised
of {\bf Baryonic matter} and, about six times as much, {\bf Nonbaryonic dark matter}.
The nonbaryonic matter's composition is unknown but can comprise particles
with any mass between one billionth of the proton mass to one millionth of the
solar mass, a range of 60 orders of magnitude. Popular candidates
arising from extension of the standard model
are the neutralino of supersymmetry and the axion arising 
from the strong CP problem. The neutralino is one example of a
WIMP or Weakly Interacting Massive Particle.
The dark matter sector may be richer than
suggested by a single constituent
and equally as complex as the baryonic matter sector
but as a first hypothesis it is natural to assume a single dark
matter constituent. 

{\bf Doppler effect}

The frequency of a wave which may be sound or light is effected by
the relative motion of the source and observer. This
effect was first discovered by Doppler in 1845. In a dramatic
demonstration of his discovery, he arranged for trumpeters
to play on an open wagon of a moving train and for
musicians with perfect pitch to listen to the change
of pitch as the train passed. This physics, applied to electromagnetic
waves, played a central role 80 years later in the discovery of the expansion of the universe.

{\bf Earth}

Third nearest planet to the Sun and the only known location
of life. In the universe there must be either
only one or many such locations: two would be much  harder
to comprehend than one. The reason is this. The 
Milky Way contains one hundred billion stars
similar to our Sun; the visible universe contains one hundred
billion galaxies similar to our Milky Way so there are some
ten billion trillion stars. Over one hundred planets have been
discovered near relatively nearby stars so there are presumably
some hundred billion trillion planets. The probability
of life originating on a given planet is more or less
than one in one hundred billion trillion. If it is much less,
we are likely alone; if the probability is much larger
there should be many life bearing planets. 
The probalility of life formation on a given planet
is unlikely to be 
close to the reciprocal of the total number of planets
in the visible universe.

{\bf Electromagnetic radiation}

Oscillating and propagating electric and magnetic fields. Examples are
X rays, radio waves, visible light. The cosmic microwave background
is an important example, originating from the {\bf surface of last
scatter}. At the present temperature of 2.725 degrees the peak
frequency is about one millimeter and therefore in
the microwave region.

{\bf Friedmann equation}

Mathematical equation governing the time dependence of the {\bf Scale factor}
derived by Friedmann in 1922. It was independently derived by Lema\^{i}tre
in 1927 and is in some books called the Friedmann-Lema\^{i}tre
equation. In this book for simplicity it is called the Friedmann equation. 

{\bf General Relativity}

Theory relating gravitation to the geometry of space introduced by
Einstein in 1915 after concentrating for ten years on how to generalize
special relativity to the case of accelerating coordinate systems. An
important intuition was that a falling person does not
experience the force of gravity, implying the equality of gravitational
and inertial mass, as expressed by the equivalence principle.
The theory explained the discrepancy in the perihelion
of Mercury when calculated from Newton's law and made two
predictions: the bending of light confirmed in the solar eclipse
expedition of 1919 and the red-shift of ``falling'' light
confirmed by Pound and Rebka in 1960.

{\bf Helium}

The second lightest element after hydrogen.
Also, by far the second most common atom after hydrogen.
The fact that the abundances are about 23\% by mass
of helium and 77\% by mass of hydrogen strongly
support Big Bang nucleosynthesis; the other evidences
for the Big Bang are the discovery of the cosmic
expansion and of the cosmic microwave backgound.

{\bf Homogeneous}

Uniformly distributed or the same at every position.
At large distances, the universe does appear
to become more and more homogeneous and so
this, together with isotropy, are the traditional assumptions
made about the universe in theoretical cosmology.

{\bf Hubble Law}

Recession velocity is proportional to distance.
This was established by Edwin Hubble in 1929
and is the most important discovery
made in cosmology before the accelerated rate of expansion
was discovered in 1998.. Hubble's original data was
so inaccurate that it's even more impressive
that he drew the correct conclusion. This expansion
of the universe convinced Einstein, as he expressed it
to Hubble on a visit to California in 1931, that his 1917
paper on cosmology had contained the ``blunder''
of introducing a cosmological constant.

{\bf Hubble parameter}

Ratio of recession speed to distance. Its units are
kilometers per second per megaparsec. In these units
the value is now known to be within a few percent of 70.
Historically, its value was controversial because of
uncertainties in distance measurement. For several decades
values as different as 50 and 100, both with quite small
errors, were forcefully
defended by various groups. Hubble's original
value was even worse, close to 500, back in 1929. 

{\bf Hubble Space Telescope}

An orbiting instrument with a 2.4 meter primary mirror
deployed by NASA astronauts in 1990. 
The HST has been one of the most successful scientific 
instruments, having made a series of important discoveries.

{\bf Hydrogen}

The lightest element whose atomic nucleus contains one proton.
The atome contains also one electron. This is the simplest
atomic system and so formed the first target of quantum
mechanics which explaines its detailed spectrum.
About 75\% of the atoms in the universe are hydrogen,
while 25
``metallic'' atoms is negligible.

{\bf Inflation}

Period of very rapid expansion during the first
trillionth of a second after the Big Bang.
Appended to the Big Bang, inflation solves two
severe fine-tuning issues: the horizon and flatness problems,
as discussed in Chapter Two of this book.

{\bf Inflaton}

Hypothesized particle underlying {\bf inflation}.
Inflation was motivated by the scalar potentials and
phase transitions occurring in grand
unified theories of elementary particles. The
inflationary era is expected to originate from
the variation in the value of a scalar field named the
inflaton.  Different assumptions about the inflaton
potential leads to alternative versions of
inflation theory which observations may discriminate.

{\bf Isotropic}

The same in all directions, also known as rotational invariance.
The CMB, in particular, is isotropic to an accuracy of
one part in one hundred thousand and this represents the 
horizon problem for the unadorned Big Bang theory. The
surface of last scatter from which the CMB photons
originate contains about fourty thousand regions 
never causally connected so why should their
temperature be the same to this accuracy? The
isotropy, by itself, strongly suggests that something
like inflation occurred in the early universe.
 
{\bf LHC}

Large Hadron Collider at {\bf CERN}.
It will start test running in 2007 and collide protons in 2008 at
a center-of-mass energy of 14 TeV,
seven times the previous highest energy. In this
new energy regime, it is expected to discover the Higgs boson and
hopefully other new particles and forces which will
shed light on some of the mysteries of the standard model.

{\bf Mercury}

Nearest planet to the Sun. The known discrepancy of the
orbit of Mercury from the Newtonian prediction
was crucial in the development of general relativity.
The perihelion is the closest approach of a planet to the Sun.
It was known already before Einstein's work that there
was an anomalous precession in the perihelion
of Mercury of 43 seconds of arc per century. General
relativity was able to explain this discrepancy accurately
and to make two other predictions as classic tests:
the bending of light by the Sun, confirmed in 1919
an expedition to view the Hyades cluster
during a six-minute-long total solar eclipse from
northern Brazil and Principe, a small island West of equatorial Africa,
in 1919 and the red-shift of ``falling'' photons, first
confirmed by Pound and Rebka
at Harvard in 1960. When the eclipse results
were announced in November 1919 the exceptional media reaction
rendered Einstein a famous personality for
the rest of his life. When the expedition's leader,
Eddington, was asked at that time by a journalist
if it was true only three people understood general relativity,
his self-revealing reply was: "Who is the third?"

{\bf Milky Way}

Name of the galaxy containing the Earth. It is just one
of at least one hundred billion galaxies
in the expanding universe. The Milky Way contains
some hundred billion stars comparable to our Sun. The
Sun is thirty thousand light-years from the core
of the Milky Way whose luminous matter stretches
out to one hundred thousand light-years in spiral arms.  
From outside, the Milky Way would presumably look
like one of the many other observed spiral galaxies
although such a picture is not likely anytime soon.
The Milky Way contains about ten times as much
mass in dark matter stretching a few hundred thousand
light-years from the core, far enough
to mingle with the corresponding
dark matter halo of neighbouring Andromeda.

{\bf Naturalness}

Criterion that a theory should not contain
unexplained large ratios of numbers. If there exists
an unexplained large dimensionless ratio, it is
often called a hierarchy. As much has been written
about naturalness and hierarchy as any other topic in 
particle theory. It explains the popularity
of low-scale supersymmetry despite no experimental
evidence. Its importance was stressed strongly in 1971 by 
the eminent theorist K.G. Wilson who, however, much later in 2004
called his assertion a ``blunder'' when he
emphasized instead that very 
large ratios can and do occur naturally in Nature.

{\bf Neutrino}

Elementary particle which experiences only the weak interaction.
Until 1998, the neutrino mass could have been zero but since then we
know that flavors (e, $\mu, \tau$) of neutrino oscillate into each
other which requires that there are non-zero mass differences.
All the three masses must be  small, less than one millionth of the electron
mass. The mixing angles, on the other hand, are large, near
maximal for two angles out of three; this is different from quarks
where all three mixing angles are small. The non-zero mass of neutrinos increases
the number of free parameters in the standard model from 19 to 28.

{\bf Nonbaryonic Dark Matter}

23\% of the energy of the universe. It clumps like baryonic matter
but non-luminous. The consitutents of dark matter are completely unknown and
can have mass ranging anywhere from a billionth of a proton mass
to a millionth of the solar mass, a range of 60 orders of magnitude.
Popular candidates arising from motivated extension of the standard model
are the neutralino particle of supersymmetry and the axion associated
with solution of the string CP problem. The neutralino is expected
with a mass of about 100 times the proton mass and exemplifies a
class of dark matter particles called WIMPs (Weakly Interacting
Massive Particles). The axion is expected to be near the light
end, between one billionth and one millionth of the proton mass.  
The dark matter could be much more complicated and contain
a wide variety of constituents, just as does the luminous
baryonic matter.

{\bf Nontrivial topology}

An idea that space has special properties at the largest distance scales.
There is no evidence for this but the accurate data on the CMB
will allow detailed searches for unexpected properties such
as ``circles in the sky'' where circles in opposite directions
exhibit correlations which reflect a periodicity of space.
So far, no such correlations have been found but if they
were, it would change significantly our notion of space dimensions.

{\bf Photon}

Massless elementary particle which is the smallest unit of 
{\bf electromagnetic radiation}. The photon was used in
his explantion of the photoelectric effect by Einstein
in 1905 but it was not universally accepted until 1923
with the demonstration of the Compton effect in which
light must be treated as particulate. Two prominent skeptics
of the interim period were Planck himself in 1913 who,
in nominating Einstein to the Prussian Academy
asked the committee to overlook the work on photons as
an example where Einstein ``may have gone overboard in his
specualtions''; and Bohr, in his 1922 Nobel lecture,
said that photons are ``not able to shed light on the nature of
radiation''.

{\bf Principia}

Exceptional book by Isaac Newton published in 1687, rightly
regarded as the beginning of modern theoretical physics. For the next two
hundred years this book, now of only historical interest, had huge influence.
It contains in particular the three laws of motion and the universal
law of gravity. Although the author invented calculus, he remarkably
avoided using it in the book, although this made many of his proofs
more complicated. Over three hundred years later, non-calculus
courses in introductory physics are still offered by many universities.

{\bf Quantum Mechanics}

Theory invented in 1925
which explains spectra and stability of atoms. According to
classical theory an electron circulating around a nucleus
would rapidly lose energy by radiation. In quantum mechanics, a basic
assumption is that the energy is quantized and cannot be
lower than that of a ground state which is the stable situation
of an atom. Quantum mechanics explained the spectra of atomic radiation
as well as many other hitherto profound
mysteries such as the distinction between conductors
and insulators of electricity. It introduces a
fundamental constant of Nature which was
discovered phenomenologically in 1900 by Planck.

{\bf Recombination}

Cosmological era when charged particles bind to form neutral atoms
and universe becomes transparent to {\bf electromagnetic radiation}.
This creates the {\bf surface of last scatter} from which
the CMB radiates. ``Recombination'' is a misnomer because the
electrons and protons were never combined.

{\bf Scale factor}

Measures distance between galaxies and depends on time 
according to the {\bf Friedmann equation}. A useful analogy is the surface
of a balloon with spots painted on its surface. As the balloon inflated the 
spots separate from each other. The galaxies separate similarly as the universe expands.

{\bf Scientific Method}

Systematic making of experimental observations, framing hypotheses and theories
which make predictions followed by more experimental tests.
The ancient Greeks proceeded in a different fashion by 
assuming that human thought alone could discover the laws of
physics. It required the intellects of people like Galileo 
and Newton, after a remarkable two thousand years of dark ages,
to identify experiments and observations as the 
driving force of physics with human thought secondary.
They realized that all humans are less smart than Nature.

{\bf Solar System}

Our Sun and its eight planets including Mercury, Venus, Earth,
Mars, Jupiter, Saturn, Uranus, and Neptune. General relativity
has been tested with good precision on the length scales
which characterize the solar system. The visible universe has
a size 14 orders of magnitude bigger and
general relativity is not tested with comparable precision at such
length scales.

{\bf Special Relativity}

Theory which accommodates motion comparable to the speed of light,
proposed by Einstein in 1905 on the basis of a symmetry of Maxwell's
equations which he generalized also to classical mechanics, and
all of theoretical physics. It placed time on a similar
footong to the three space coordinates and led to the notion
of four-dimensional spacetime. The predictions of special relativity
for motion at ultra-high speed have been confirmed
to impressive accuracy.

{\bf Steady-state theory}

A once popular, now discredited, alternative to the {\bf Big Bang},
in which the universe is quasi-static. It postulated that the dilution of 
matter caused by the universe's expansion was compensated by the 
continuous creation of new matter. The discovery of the CMB in 1965
laid the steady-state theory to rest.

{\bf String theory}

An elegant and consistent mathematical framework proposed as a marriage
of quantum mechanics and general relativity, based on
the premise that the fundamental entities are not point particles but
extended objects, the simplest being one-dimensional strings. 
Experimental effects of quantum gravity
are probably so extremely small that they will not
be detected in the foreseeable future so it will be impracticable to
determine whether string theory is the correct theory
of quantum gravity. String theory has a dual description
by gauge field theories and hence provides a source of interesting ideas of
how to extend the established standard model of particle phenomenology.
Independently of whether string theory is the correct
theory of gravity (a question unlikely to be answered anytime soon), 
it is the only known consistent extension of
quantum field theory from point particles to extended objects
As such it provides an important insipration for model-building 
in the simpler, and more realistically testable, gauge field theories.

{\bf Sun}

By far the nearest and most understood star. Its present age
is about five billion years and it is expected to run out
of hydrogen fuel in another similar time. The next nearest
star is remarkably a million times further away and much harder to
study. Our understanding of the Sun's evolution over
its five-billion year history is remarkably good and
enshrined in a theory called the Standard Solar Model.

{\bf Surface of last scatter}

An opaque wall corresponding to {\bf recombination} where
photons of {\bf cosmic microwave background} originate.
The photons we observe in the CMB have travelled uninterrupted
through the universe for over 13 billion years from this
surface which makes an effectively-opaque wall behind which
(or earlier than which) electromagnetic radiation cannot
penetrate.

{\bf T-duality}

Mathematical symmetry of string theory which relates very
small and very large distance scales. In application of
string theory to cosmology, T-duality plays a central role
and leads to the suggestion that there is a maximum
temperature in the past and that before that there
might exist a T-dual universe in which distance scales
were interchanged between the very large and very small.
This is difficult to visualize but follows
simply from the string theory equations.

{\bf Turnaround}

The transition from expansion and contraction phases in
a cyclic cosmology.

{\bf Venus}

Second nearest planet, after Mercury, to the Sun. The two
inner planets provide some of the best checks of general
relativity theory.

{\bf WIMP}

Weakly Interacting Massive Particle, a candidate for the
cold dark matter particle. One example is the neutralino
of supersymmetric theories but the WIMP is a more general
concept of a particle which might be discovered at the LHC.

{\bf WMAP}

Wilkinson Microwave Anisotropy Probe, whose first data
was released in February 2003 and initiated ``Precision Cosmology''
by virtue of analysis of its data leading to values of the cosmic
parameters with unprecedentedly small errors. 

\bigskip
\bigskip
\bigskip
\bigskip
\bigskip

\end{document}